\documentclass[fleqn,usenatbib]{mnras}

\usepackage{newtxtext,newtxmath}

\usepackage[T1]{fontenc}
\usepackage{float}
\usepackage{physics}
\usepackage{subfiles}
\usepackage{comment}
\usepackage{booktabs}
\usepackage{graphics}
\usepackage{comment}
\usepackage{graphicx}
\usepackage{hyperref}
\usepackage{xcolor}
\usepackage{lipsum} 
\usepackage{tabularx}
\usepackage{graphicx}
\usepackage{orcidlink}
\DeclareRobustCommand{\VAN}[3]{#2}
\let\VANthebibliography\thebibliography
\def\thebibliography{\DeclareRobustCommand{\VAN}[3]{##3}\VANthebibliography}

\usepackage{graphicx}	
\usepackage{amsmath}	
\usepackage{xcolor}

\definecolor{green_cust}{HTML}{00FF00}

\definecolor{blue_cust}{HTML}{0066ff}

\definecolor{gold_cust}{HTML}{FFBF00}

\definecolor{purple_cust}{HTML}{800080}

\definecolor{pink_cust}{HTML}{DC267F}

\definecolor{perfect_green}{HTML}{4FBF26}

\definecolor{crimson}{HTML}{DC143C}

   \title[Collisional and Radiative Data for Te I-III]{Collisional and Radiative Data for Tellurium ions in Kilonovae modelling and Laboratory Benchmarks}

\author[L.P. Mulholland et al.]{
L. P. Mulholland,\orcidlink{0009-0003-2668-5589}$^{1}$\thanks{E-mail: lmulholland25@qub.ac.uk}
F. McNeill,\orcidlink{0009-0001-9528-7475}$^{1}$
S. A. Sim,\orcidlink{0000-0002-9774-1192}$^{1}$
C. P. Ballance \orcidlink{0000-0003-1693-1793}$^{1}$
and
C. A. Ramsbottom \orcidlink{0000-0003-1579-8556}$^{1}$
\\
$^{1}$Astrophysics Research Centre, School of Mathematics \& Physics, Queens University Belfast, BT7 1NN, Northern Ireland.
}

\date{Accepted XXX. Received YYY; in original form ZZZ}

\pubyear{\the\year{}}

\begin{document}
\label{firstpage}
\pagerange{\pageref{firstpage}--\pageref{lastpage}}
\maketitle

\begin{abstract}
    Tellurium is a primary candidate for the identification of the 2.1 $\mu$m emission line in kilonovae (KNe)  spectra AT2017gfo and GRB230307A. Despite this, there is currently an insufficient amount of atomic data available for this species. We calculate the required atomic structure and collisional data, particularly the data required for accurate Non-Local-Thermodynamic-Equilibrium (NLTE) modelling of the low temperatures and densities in KNe. We use a Multi-Configurational-Dirac-Hartree-Fock method to produce optimised one-electron orbitals for Te {\sc i}-{\sc iii}. 
   As a result energy levels and Einstein A-coefficients for Te {\sc i}-{\sc iii} have been calculated. These orbitals are then employed within Dirac $R$-matrix collision calculations to provide electron-impact-excitation collision strengths that were subsequently averaged according to a thermal Maxwellian distribution.  Subsequent \textsc{tardis} simulations using this new atomic data reveal no significant changes to the synthetic spectra due to the very minor contribution of Te at early epochs. NLTE simulations with the ColRadPy package reveal optically thin spectra consistent with the increasing prominence of the Te {\sc iii} 2.1 $\mu$m line as the KNe ejecta cools.  This is reinforced by the estimation of luminosities at nebular KNe conditions. New line ratios for both observation and laboratory benchmarks of the atomic data are proposed. 
\end{abstract}


\section{Introduction}

Following the observation of GW170817 \citep{abbott2017gw170817} along with its electromagnetic counterpart AT2017gfo, a large effort has been put towards identifying the elements responsible for the features in its spectrum \citep{smartt2017kilonova,Hotokezaka2022WSe,hotokezaka2023tellurium,gillanders2024modelling}.  Previously Sr {\sc ii} and Y {\sc ii} lines have been identified \citep{watson2019identification,sneppen2023discovery,domoto2021} and these identifications have recently been supplemented by  \cite{Mulholland24} with newly computed radiative and collisional atomic data. 
Tellurium lies at the second peak of the $r$-process, making it an attractive candidate for potential identification in kilonova spectra.
Of particular interest is the broad emission feature observed at approximately $2.1 \mu \text{m}$. This line has been suggested  \citep{hotokezaka2023tellurium} to be due to a forbidden transition between fine-structure levels of the ground state of Te {\sc iii}. This was reinforced by the work of  \citet{gillanders2024modelling}. In both cases, it was pointed out that the neutral case Te {\sc i} has two overlapping lines at around this wavelength, suggesting a possible blend or alternative identification.  Additionally, Te {\sc i} was also suggested as a possible contribution to the observed absorption at rest wavelength of approximately $1.0$ $\mu$m in AT2017gfo, (see \citealt{smartt2017kilonova} but also discussion by \citealt{watson2019identification}), and forbidden lines of Te {\sc i and \sc ii} have been suggested as candidate identifications for emission features by \cite{gillanders2024modelling}. Recently, the James Webb Space Telescope observed GRB 230307A \citep{2023GCN.33405....1F,levan2024heavy}. The resulting spectrum is thought to contain emission from the kilonova associated with a merger, and exhibits an emission feature at a wavelength close to $2.1 \mu \text{m}$, which has also been identified with Te {\sc iii} \citep{gillanders2023heavy,levan2024heavy}.  Beyond KNe modelling, Te {\sc iii} has been previously identified in planetary nebulae NGC 7027 and IC 418 \citep{madonna2018neutron} and Te {\sc i} has been identified in the metal-poor stars BD +17 3248, HD 108317, and HD 128279 \citep{roederer2012detection}, where the inferred abundances rely on high-quality atomic data.

Recent works have suggested a need to move away from the approximation of local thermodynamic equilibrium (LTE), particularly for accurate representation of the relatively low density plasmas present in KNe \citep{mccann2022atomic}. There have been relatively few works on the NLTE modelling of KNe \citet{PognanNLTE,hotokezaka2023tellurium,Tarumi2023}. With a general lack of  atomic data required for NLTE modelling, the semi-empirical approximations of \citet{van1962rate} or \citet{Axelrod1980} are sometimes used. It has been previously shown by \citet{bromley2023electron}, \citet{mccann2022atomic} and more recently \citet{Mulholland24}, that such approximations are inadequate and inaccurately model the resulting populations. 
For this purpose, we aim in this publication to expand on the public atomic data sets by presenting electron-impact-excitation and emission rates for Te {\sc i-iii} required for NLTE modelling.

It is therefore clear that the near neutral ion stages of Te are of particular interest in current astrophysical research and there is a need for atomic data for both radiative and collisional processes, as required for the study of low-density NLTE plasmas. 

The neutral species has had little attention in recent years with the spectral measurements of relevant energy levels from \citet{morillon1975observation} remaining the primary dataset. A small number of transition probabilities have been published by \citet{ubelis1983transition} with six lines in the NIST ASD \citep{nist}. The most recent measurements of atomic data for Te {\sc ii} and {\sc iii} have been the experimental studies from \citet{tauheed2011revision} with revised identifications from the theoretical work of \citet{zhang2013transition}, who also publish a limited set of oscillator strengths for the electric-dipole transitions for Te {\sc ii} and Te {\sc iii}. Theoretical calculations for the $N=51$ isoelectronic sequence, including Te {\sc ii},  have been performed by \citet{radvziute2023theoretical}. Additionally, semi-relativistic collision strengths for ground-state transitions of Te {\sc iii} were calculated by \citet{madonna2018neutron}.

Here we aim to address the need for tellurium atomic data and expand on the public atomic data sets by presenting electron-impact-excitation and emission rates for Te {\sc i-iii}, as required for NLTE modelling.
The remainder of this paper is structured as follows. In Section \ref{sec:atomic_structure} we briefly describe the mode of operation of our chosen atomic structure implementation, namely {\sc grasp$^0$}, before presenting our calculated atomic structures and Einstein A-coefficients for the first three ion stages of Tellurium ($Z = 52$). In Section \ref{sec:electron_impact_excitation} we describe the relativistic electron-impact-excitation calculations and show the collision strengths and Maxwellian-averaged collision strengths for some representative transitions of interest. Collisional radiative modelling via ColRadPy is presented in Section \ref{sec:colrad} where potential diagnostics are identified and synthetic spectra are shown. In Section \ref{sec:tardis} we perform 1D LTE synthetic spectral modelling using the \textsc{tardis} radiative transfer code for early phases of the event AT2017gfo, and perform a differential comparison of the newly calculated atomic data and the literature values.  Finally we conclude with a summary and outlook in Section \ref{sec:conclusions}.

\section{Atomic Structure}\label{sec:atomic_structure}

Optimised orbitals for Te {\sc i}-{\sc iii} were generated using the {\sc grasp$^0$} package \citep{Grant80,Dyall1989}, where an Extended-Average-Level (EAL) method provides orbital optimisation over
all included configurations.
This procedure weights the diagonal elements of the Dirac-Coulomb Hamiltonian (in atomic units),
\begin{equation}
H_{DC} = \sum_i \bigg(c \boldsymbol{\alpha}\cdot\boldsymbol{p}_i + (\beta -I_4)c^2 -\frac{Z}{r_i}\bigg) + \sum_{i>j}\frac{1}{r_{ij}},
\end{equation}
according to the statistical weight of the corresponding configuration-state-function (CSF). Here, $\boldsymbol{\alpha}$ and $\beta$ are the set of four Dirac-matrices, $I_4$ is the 4 $\times $ 4 identity matrix, $c$ is the speed of light, $r_i$ is the radial position of electron $i$, $r_{ij}$ is the inter-electron distance and $Z$ is the nuclear charge. A Multi-Configurational-Dirac-Hartree-Fock (MCDHF) variational method is used to optimise the orbitals, which are subsequently employed in the electron-impact collisional calculation. Typically configuration choice is assisted by initially comparing the calculated energy levels and Einstein A-coefficients  $A_{jk}$ for each system with available theoretical or  experimental data. The size of the calculation should be kept relatively small, as the subsequent scattering calculation grows in complexity to the third power with the number of included configurations. A balance must be struck between accuracy and size. The values of $A_{jk}$ presented here are  adjusted by shifting the wavelengths to spectroscopic values via,
\begin{equation}
    A_{\text{shifted}} = \Big(\frac{\lambda_{\text{calc}}}{\lambda_{\text{expt}}}\Big)^3 A_{\text{calc}},
\end{equation}
for E1 and M1 transitions with similar relations for higher order transitions, where the wavelength ratio increases to a power 5 for quadrupoles with an associated impact on the A-value.

\begin{table}
    \centering
    \begin{tabular}{c l }
    \hline \\
    Te {\sc i} Model   & 5s$^2$5p$^4$; 5s5p$^4$\{5d,6s,6d,7s\};           \\
    - 29 CSF       & 5s$^2$5p$^3$\{6s,6p,5d,6d,7s,7p\};            \\
                       & 5s$\hphantom{^2}$5p$^3$\{6s$^2$,6p$^2$\}; 5s5p5d$^4$          \\
                       & 5s$^2$5p$^2$\{5d$^2$,6s$^2$,6s6p,6d$^2$,7s$^2$,7p$^2$\};   \\ 
                       & 4d$^{10}$ \{5p$^6$,5p$^5$5d,5p$^4$6s$^2$,5p$^4$5d$^2$,5p$^3$6s6p$^2$\};   \\
                       & 5s$^2$5p\{5d$^3$, 6d$^3$,6s6p$^2$\}; 5s5p$^5$.   \vspace{4mm}\\
    Te {\sc ii} Model  & 5s$^2$ 5p$^{3}$; 5s\{5p$^{4}$,6p$^{4}$ \}           \\
     - 27 CSF      & 5s$^2$ 5p$^{2}$ \{6s,6p,5d,6d,7s,7p\};            \\
                       & 5s$^2$ 5p$\hphantom{^{1}}$ \{6s6p,5d$^{2}$,6d$^{2}$,6p$^{2}$,7p$^{2}$\};          \\
                       & 5s$^2$ \{5d$^{3}$,6p$^{3}$,7p$^{3}$\};           \\
                       & 5s$\hphantom{^2}$ 5p$^{2}$ \{6s$^{2}$,6p$^{2}$,7p$^{2}$\};   \\ 
                       & 5s$\hphantom{^2}$ 5p$^{3}$\{5d,6s\};    \\
                       & 5s$\hphantom{^2}$ 6p$^{2}$7p$^{2}$; 4d$^{10}$ 5p$^{5}$;   \\
                       & 4d$^{9}$ 5s$^2$ 5p$^{4}$;   \\
                       & 4d$^{8}$ \{5s$^2$ 5p$^{5}$,5s 5p$^{6}$\}.   \vspace{4mm} \\
    Te {\sc iii} Model & 5s$^2$\{5p$^2$,5d$^2$,6s$^2$,6p$^2$,6d$^2$\};           \\
     - 24 CSF      & 5s$^2$5p\{5d,6s,6p,6d\};           \\
                       & 5s$\hphantom{^2}$\{5p$^3$,5d$^3$,6p$^3$,6d$^3$,5p$^2$5d,5p$^2$6d,5p$^2$6s\}         \\
                       & 4d$^{10}$5p$^3$\{6s,5d \}  \\
                       & 4d$^{10}$5p$^2$\{6p$^2$,5d$^2$,6s$^2$ \}  \\
                       & 4d$^{10}$ \{5p$^4$,6p$^4$,5p5d$^3$ \}. \vspace{1mm}  \\

    \hline\\
    \end{tabular}
    \caption{The configurations included in the wavefunction expansion for the three structure calculations.}
    \label{tab:csfs}
    \end{table}

In calculating the atomic structures, the non-relativistic valence configurations listed in Table \ref{tab:csfs} were included in each of the calculations. The configurations were chosen to optimise agreement with the experimentally measured levels and available oscillator strengths. In all three calculations, the initial guess for the one-electron orbitals were hydrogenic, but were subsequently refined according to a self consistent field procedure. The remainder of this section will compare the theoretical energies calculated here with experiment and present transition probabilities for transitions of interest with comparisons made where possible. The calculated $A_{ji}$ will be made available in the standard {\sc adf04} format at \cite{openadas_site} (along with the Maxwellian Averaged Collision strengths to be discussed in Section \ref{sec:electron_impact_excitation}).

With the relatively low temperatures present in KNe, forbidden transitions within the ground state of Te ions are expected to be particularly important. For such transitions, accuracy is hard to assess given the difficulty of their experimental measurement as well as the requirement of good convergence for the ground state in such Configuration Interaction (CI) calculations. In this regard we will for each ion compare our calculated data for forbidden transitions with those computed by \cite{biemont1995forbidden} who detail ground state transition data for the 5p$^k$ set of ions. Additionally the scarcity of published strong-dipole transitions motivates us to compare with other theoretical datasets where available \citep{zhang2013transition,biemont1995forbidden,madonna2018neutron}. While such transitions may not be useful directly for KNe modelling, they provide useful indications of the quality of the atomic structure model itself - in particular the atomic orbitals that are carried forward to the scattering calculations. Additionally, they are included in the published data-sets for completeness of the collisional-radiative model and for use in wider applications.

\subsection{Te {\sc i}}

The MCDHF iterative algorithm as implemented within {\sc grasp$^0$} may have difficulty converging variationally determined orbitals with high $n$ and $l$ for complex neutral systems. The 29 non-relativistic valence configurations in Table \ref{tab:csfs} were used to optimise the structure. This resulted in a large 1653 fine-structure level calculation, the first 30 of which are compared with those listed on the NIST database \citep{nist} in Table \ref{table:energy_levels_neutral}. Generally good agreement is obtained. It can be seen that the energies of the $6$s levels are slightly overestimated, with the energies of the other states generally underestimated. The average-absolute-percentage error lies at 5.3\%. The even and odd states are approximately equally well represented with an average absolute error of 5.2\% and 5.5\% respectively. These are shifted, where available, to spectroscopic values in calculating the A-values and collision strengths.

\begin{table}
    \centering
    \begin{tabularx}{0.475\textwidth}{lllllrr}
        \toprule
        Index & CSF & Level  & Expt & {\sc grasp$^0$} & $\Delta E$ & \% \\ 
        \midrule
        1  &5s$^2$ 5p$^4$    & $^3$P$_2$             & 0.0000 & 0.0000 &  0.0000 &   0.00 \\ 
        2  &5s$^2$ 5p$^4$    & $^3$P$_0$             & 0.0429 & 0.0399 & -0.0030 &  -6.93 \\ 
        3  &5s$^2$ 5p$^4$    & $^3$P$_1$             & 0.0433 & 0.0380 & -0.0053 & -12.31 \\ 
        4  &5s$^2$ 5p$^4$    & $^1$D$_2$             & 0.0962 & 0.1115 &  0.0153 &  15.94 \\ 
        5  &5s$^2$ 5p$^4$    & $^1$S$_0$             & 0.2114 & 0.2169 &  0.0055 &   2.59 \\ 
        6  &5s$^2$ 5p$^3$ 6s & $^5$S$^{\rm{o}}_2$    & 0.4033 & 0.4049 &  0.0017 &   0.41 \\ 
        7  &5s$^2$ 5p$^3$ 6s & $^3$S$^{\rm{o}}_1$    & 0.4251 & 0.4355 &  0.0104 &   2.44 \\ 
        8  &5s$^2$ 5p$^3$ 6p & $^5$P$_1$             & 0.4935 & 0.4685 & -0.0250 &  -5.07 \\ 
        9  &5s$^2$ 5p$^3$ 6p & $^5$P$_2$             & 0.4939 & 0.4690 & -0.0249 &  -5.04 \\ 
        10 &5s$^2$ 5p$^3$ 6p & $^5$P$_3$             & 0.4970 & 0.4710 & -0.0259 &  -5.22 \\ 
        11 &5s$^2$ 5p$^3$ 6s & $^3$D$^{\rm{o}}_1$    & 0.4983 & 0.5483 &  0.0500 &  10.03 \\ 
        12 &5s$^2$ 5p$^3$ 6s & $^3$D$^{\rm{o}}_2$    & 0.5001 & 0.5488 &  0.0487 &   9.75 \\ 
        13 &5s$^2$ 5p$^3$ 6p & $^3$P$_1$             & 0.5044 & 0.4853 & -0.0192 &  -3.80 \\ 
        14 &5s$^2$ 5p$^3$ 6p & $^3$P$_2$             & 0.5073 & 0.4869 & -0.0203 &  -4.01 \\ 
        15 &5s$^2$ 5p$^3$ 5d & $^5$D$^{\rm{o}}_3$    & 0.5074 & 0.4875 & -0.0199 &  -3.92 \\ 
        16 &5s$^2$ 5p$^3$ 6p & $^3$P$_0$             & 0.5086 & 0.4877 & -0.0208 &  -4.09 \\ 
        17 &5s$^2$ 5p$^3$ 5d & $^5$D$^{\rm{o}}_4$    & 0.5086 & 0.4873 & -0.0213 &  -4.19 \\ 
        18 &5s$^2$ 5p$^3$ 5d & $^5$D$^{\rm{o}}_2$    & 0.5086 & 0.4875 & -0.0211 &  -4.16 \\ 
        19 &5s$^2$ 5p$^3$ 5d & $^5$D$^{\rm{o}}_0$    & 0.5087 & 0.4878 & -0.0209 &  -4.12 \\ 
        20 &5s$^2$ 5p$^3$ 5d & $^5$D$^{\rm{o}}_1$    & 0.5090 & 0.4876 & -0.0213 &  -4.19 \\ 
        21 &5s$^2$ 5p$^3$ 6s & $^3$D$^{\rm{o}}_3$    & 0.5180 & 0.5628 &  0.0449 &   8.66 \\ 
        22 &5s$^2$ 5p$^3$ 6s & $^1$D$^{\rm{o}}_2$    & 0.5205 & 0.5726 &  0.0521 &  10.02 \\ 
        23 &5s$^2$ 5p$^3$ 5d & $^3$D$^{\rm{o}}_2$    & 0.5339 & 0.5019 & -0.0321 &  -6.01 \\ 
        24 &5s$^2$ 5p$^3$ 5d & $^3$D$^{\rm{o}}_1$    & 0.5353 & 0.5027 & -0.0326 &  -6.09 \\ 
        25 &5s$^2$ 5p$^3$ 5d & $^3$D$^{\rm{o}}_3$    & 0.5361 & 0.5042 & -0.0319 &  -5.95 \\ 
        26 &5s$^2$ 5p$^3$ 7s & $^5$S$^{\rm{o}}_2$    & 0.5525 & 0.5224 & -0.0301 &  -5.45 \\ 
        27 &5s$^2$ 5p$^3$ 7s & $^3$S$^{\rm{o}}_1$    & 0.5571 & 0.5404 & -0.0167 &  -2.99 \\ 
        28 &5s$^2$ 5p$^3$ 5d & $^3$P$^{\rm{o}}_2$    & 0.5768 & 0.5931 &  0.0163 &   2.82 \\ 
        29 &5s$^2$ 5p$^3$ 7p & $^5$P$_1$             & 0.5792 & 0.5662 & -0.0130 &  -2.24 \\ 
        30 &5s$^2$ 5p$^3$ 7p & $^5$P$_2$             & 0.5797 & 0.5660 & -0.0137 &  -2.36 \\ 
        \bottomrule
    \end{tabularx}

    \caption{The first 30 experimental energy levels (in Rydbergs, Ryd) of Te {\sc i} compared with our atomic structure model.}
    \label{table:energy_levels_neutral}
\end{table}

\begin{figure} 
\centering
    \includegraphics[width = \linewidth]{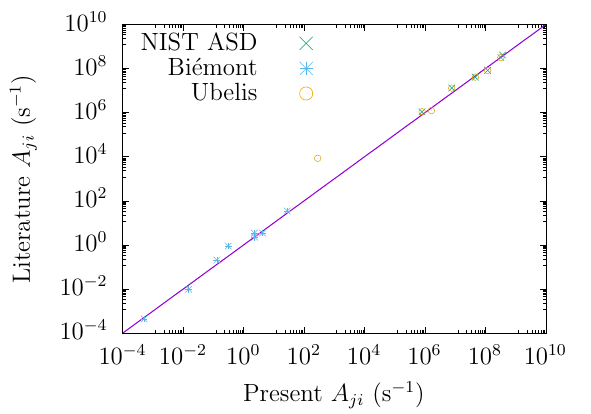}
    \caption{The present transition probabilities $A_{ji}$ for Te {\sc i} compared with the data available in the NIST ASD  \citep{nist} and the theoretical calculations of \citet{biemont1995forbidden}.}
    \label{fig:Te0+:avalues}
\end{figure}

The neutral system presents a particular gap in the literature, emphasised by the use of $\log gf = 0.0$ for certain Te {\sc i} transitions in KNe modelling \citep{smartt2017kilonova}.  We compare the calculated ground state transitions with the calculations of \citet{biemont1995forbidden} on Figure \ref{fig:Te0+:avalues}. In general, good agreement is seen in these transitions. The data direct from \citet{ubelis1983transition} is also included in this comparison as there are additional lines computed in this work that are not available in the NIST ASD. The transition probabilities published by \citet{ganas1997oscillator,ubelis1983transition} and compiled by \citet{morton2000atomic,nist} are compared (collectively labelled as NIST ASD) with calculated values in Figure \ref{fig:Te0+:avalues} where good agreement is seen.

\subsection{Te {\sc ii}}

\begin{table}
    \centering
    \begin{tabularx}{0.475\textwidth}{lllllrr}
        \toprule
        \hfil Index & \hfil CSF & \hfil Level  & \hfil Expt   & {\sc grasp$^0$} & $\Delta E$\hphantom{-} & \%\hphantom{-} \\
        \midrule
        1  & 5s$^2$5p$^3$            & $^4$S$^{\rm{o}}_{3/2}$    &0.0000  & 0.0000&	0.0000 &  0.00 \\
        2  & 5s$^2$5p$^3$            & $^2$D$^{\rm{o}}_{3/2}$    &0.0932  & 0.1204&	0.0272 & 29.23 \\
        3  & 5s$^2$5p$^3$            & $^2$D$^{\rm{o}}_{5/2}$    &0.1132  & 0.1418&	0.0286 & 25.25 \\
        4  & 5s$^2$5p$^3$            & $^2$P$^{\rm{o}}_{1/2}$    &0.1872  & 0.2035&	0.0162 & 8.67 \\
        5  & 5s$^2$5p$^3$            & $^2$P$^{\rm{o}}_{3/2}$    &0.2190  & 0.2328&	0.0138 & 6.29 \\
        6  & 5s \hspace{0.4mm}5p$^4$ & $^4$P$_{5/2}$             &0.6488  & 0.6554&	0.0066 & 1.02 \\
        7  & 5s \hspace{0.4mm}5p$^4$ & $^4$P$_{3/2}$             &0.6825  & 0.6849&	0.0024 & 0.35 \\
        8  & 5s \hspace{0.4mm}5p$^4$ & $^4$P$_{1/2}$             &0.6953  & 0.6977&	0.0024 & 0.35 \\
        9  & 5s$^2$5p$^2$6s          & $^4$P$_{1/2}$             &0.7149  & 0.7328&	0.0179 & 2.51 \\
        10 & 5s$^2$5p$^2$5d          & $^2$P$_{3/2}$             &0.7463  & 0.7770&	0.0308 & 4.12 \\
        11 & 5s$^2$5p$^2$6s          & $^4$P$_{3/2}$             &0.7540  & 0.7645&	0.0105 & 1.39 \\
        12 & 5s$^2$5p$^2$6s          & $^2$P$_{1/2}$             &0.7616  & 0.7784&	0.0168 & 2.21 \\
        13 & 5s$^2$5p$^2$5d          & $^4$F$_{5/2}$             &0.7750  & 0.8000&	0.0250 & 3.23 \\
        14 & 5s$^2$5p$^2$5d          & $^4$F$_{3/2}$             &0.7760  & 0.7968&	0.0207 & 2.67 \\
        15 & 5s$^2$5p$^2$6s          & $^4$P$_{5/2}$             &0.7800  & 0.7980&	0.0180 & 2.31 \\
        16 & 5s \hspace{0.4mm}5p$^4$ & $^2$D$_{3/2}$             &0.7906  & 0.8262&	0.0356 & 4.51 \\
        17 & 5s \hspace{0.4mm}5p$^4$ & $^2$D$_{5/2}$             &0.7965  & 0.8340&	0.0375 & 4.71 \\
        18 & 5s$^2$5p$^2$5d          & $^4$F$_{7/2}$             &0.8010  & 0.8219&	0.0209 & 2.61 \\
        19 & 5s$^2$5p$^2$5d          & $^2$P$_{1/2}$             &0.8092  & 0.8375&	0.0283 & 3.50 \\
        20 & 5s$^2$5p$^2$5d          & $^2$F$_{5/2}$             &0.8103  & 0.8463&	0.0360 & 4.44 \\
        21 & 5s$^2$5p$^2$6s          & $^2$P$_{3/2}$             &0.8107  & 0.8430&	0.0324 & 3.99 \\
        22 & 5s$^2$5p$^2$5d          & $^4$F$_{9/2}$             &0.8249  & 0.8471&	0.0222 & 2.69 \\
        23 & 5s$^2$5p$^2$5d          & $^2$F$_{7/2}$             &0.8274  & 0.8594&	0.0320 & 3.87 \\
        24 & 5s$^2$5p$^2$5d          & $^4$D$_{3/2}$             &0.8401  & 0.8724&	0.0323 & 3.84 \\
        25 & 5s$^2$5p$^2$5d          & $^4$D$_{1/2}$             &0.8447  & 0.8724&	0.0278 & 3.29 \\
        26 & 5s$^2$5p$^2$5d          & $^4$D$_{5/2}$             &0.8456  & 0.8769&	0.0314 & 3.71 \\
        27 & 5s$^2$5p$^2$6p          & $^4$D$^{\rm{o}}_{1/2}$    &0.8564  & 0.8650&	0.0086 & 1.00 \\
        28 & 5s$^2$5p$^2$6s          & $^2$D$_{5/2}$             &0.8644  & 0.9031&	0.0387 & 4.47 \\
        29 & 5s$^2$5p$^2$6s          & $^2$D$_{3/2}$             &0.8676  & 0.9047&	0.0371 & 4.28 \\
        30 & 5s$^2$5p$^2$6p          & $^4$D$^{\rm{o}}_{3/2}$    &0.8761  & 0.8869&	0.0107 & 1.23 \\
        31 & 5s$^2$5p$^2$5d          & $^4$D$_{7/2}$             &0.8797  & 0.9116&	0.0319 & 3.62 \\
        32 & 5s$^2$5p$^2$6p          & $^2$S$^{\rm{o}}_{1/2}$    &0.8910  & 0.8931&	0.0021 & 0.24 \\
        33 & 5s$^2$5p$^2$5d          & $^4$P$_{5/2}$             &0.9042  & 0.9519&	0.0477 & 5.27 \\
        34 & 5s$^2$5p$^2$6p          & $^4$S$^{\rm{o}}_{3/2}$    &0.9075  & 0.9172&	0.0098 & 1.08 \\
        35 & 5s$^2$5p$^2$6p          & $^4$D$^{\rm{o}}_{5/2}$    &0.9123  & 0.9181&	0.0058 & 0.63 \\
        36 & 5s$^2$5p$^2$5d          & $^4$P$_{3/2}$             &0.9180  & 0.9673&	0.0493 & 5.37 \\
        37 & 5s$^2$5p$^2$5d          & $^2$G$_{7/2}$             &0.9189  & 0.9765&	0.0576 & 6.27 \\
        38 & 5s$^2$5p$^2$5d          & $^2$S$_{1/2}$             &0.9210  & 0.9710&	0.0500 & 5.43 \\
        39 & 5s$^2$5p$^2$6p          & $^2$D$^{\rm{o}}_{3/2}$    &0.9224  & 0.9352&	0.0128 & 1.39 \\
        40 & 5s$^2$5p$^2$6p          & $^4$P$^{\rm{o}}_{1/2}$    &0.9238  & 0.9299&	0.0061 & 0.66 \\ 
        41 & 5s$^2$5p$^2$5d          & $^4$P$_{1/2}$             & 0.9307 & 0.9859&	0.0552 & 5.93 \\
        42 & 5s$^2$5p$^2$5d          & $^2$D$_{3/2}$             & 0.9317 & 0.9738&	0.0421 & 4.52 \\
        43 & 5s$^2$5p$^2$6p          & $^4$P$^{\rm{o}}_{5/2}$    & 0.9324 & 0.9450&	0.0126 & 1.35 \\
        44 & 5s$^2$5p$^2$5d          & $^2$D$_{5/2}$             & 0.9359 & 0.9781&	0.0422 & 4.51 \\
        45 & 5s$^2$5p$^2$6p          & $^4$D$^{\rm{o}}_{7/2}$    & 0.9396 & 0.9490&	0.0094 & 1.00 \\
        46 & 5s$^2$5p$^2$6p          & $^4$P$^{\rm{o}}_{3/2}$    & 0.9471 & 0.9585&	0.0113 & 1.20 \\
        47 & 5s$^2$5p$^2$6p          & $^2$P$^{\rm{o}}_{3/2}$    & 0.9569 & 0.9748&	0.0179 & 1.88 \\
        48 & 5s$^2$5p$^2$6p          & $^2$D$^{\rm{o}}_{5/2}$    & 0.9621 & 0.9802&	0.0180 & 1.88 \\
        49 & 5s$^2$5p$^2$6p          & $^2$P$^{\rm{o}}_{1/2}$    & 0.9670 & 0.9793&	0.0122 & 1.27 \\
        50 & 5s$^2$5p$^2$5d          & $^2$G$_{9/2}$             & \hfil -   & 0.9842 & -\hphantom{11} &  \hphantom{-0..}-\hphantom{11}\\
    \bottomrule
    \end{tabularx}

         \caption{Energy levels (in Ryd) of Te {\sc ii} compared with the experimental data \citep{nist}.}
    \label{table:energy_levels_single}
\end{table}

\begin{figure} 
\centering
    \includegraphics[width = \linewidth]{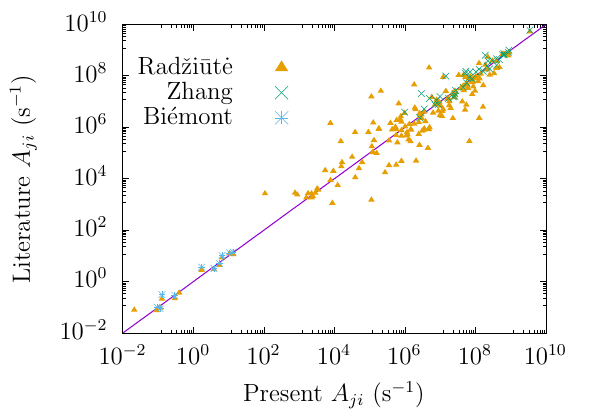}
    \caption{The present transition probabilities $A_{ji}$ for Te {\sc ii} compared with the data available from \citet{zhang2013transition} and \citet{radvziute2023theoretical} (E1 transitions) and the theoretical calculations of \citet{biemont1995forbidden} (E2 and M1 transitions) .}
    \label{fig:Te1+:avalues}
\end{figure}

For Te {\sc ii}, the  23 configurations listed in Table \ref{tab:csfs} were used to optimise the orbitals. This resulted in 506 fine-structure resolved levels. The first 50 are compared with the available literature energies available from the NIST atomic database \citep{nist} in Table \ref{table:energy_levels_single}. In agreement with \citet{zhang2013transition}, level 50, $^2$G$_{9/2}$, is predicted with no experimental verification. This fact is perhaps due to the absence of any strong dipole transitions from this state, consistent with our transition probability calculations. The average absolute difference between the two shown datasets is 4.1\%.  The largest contribution to this is the odd levels with an average error of 5.0\%, whereas the even levels contribute 3.6\%.  The first 49 levels were then shifted to the available experimental values for the transition probabilities.

For the transition probabilities, the NIST ASD \citep{nist} lists no lines for this ion. Therefore, we compare the model with the theoretical calculations of \citet{zhang2013transition,radvziute2023theoretical}.This is shown on Figure \ref{fig:Te1+:avalues} where good agreement is seen overall. Generally good agreement is seen with \cite{zhang2013transition}.  The disparity here between our data and that of \cite{radvziute2023theoretical} is larger, but there is general agreement between the two datasets.

Importantly for KNe work, the low-energy forbidden transitions are also shown on Figure \ref{fig:Te1+:avalues}. Here there is excellent agreement between the present calculations and those of \citet{biemont1995forbidden} and \citet{radvziute2023theoretical}. With good agreement between experimental energy levels and theoretical transition probabilities, there is confidence in the present target model to be used in the scattering calculation.

\subsection{Te {\sc iii}}

For Te {\sc iii}, we include the  24 non-relativistic configurations listed in Table \ref{tab:csfs}. This resulted in 644 relativistic fine-structure levels. The NIST ASD database \citep{nist} lists spectroscopic energies from \citet{moore1971}. Since this publication, new experimental energy levels have been published by \citet{joshi1992analysis,tauheed2011revision}. The identifications by \citet{tauheed2011revision} were recently revisited by \citet{zhang2013transition} and it is this dataset and identifications that we choose to compare with in Table \ref{table:energy_levels_double}. 

In agreement with \cite{tauheed2011revision,zhang2013transition,joshi1992analysis}, we predict the existence of a level 5p5d $^3$F$^{\rm{o}}_4$ (index 18) which has yet to be verified experimentally. It is noted that our corresponding oscillator strength calculations predicts no strong decays which is consistent with the lack of experimental observation.  In any case, this level lies at an energy of order 1.0 Ry making it unlikely to be excited under KNe conditions, but may prove interesting in higher energy phenomena or laboratory experiments. The average absolute error between these energy levels and our calculation is 4.1\%. The even levels have an average error of 2.9\%, while the odd levels have a larger 4.8\%. Of the 40 levels displayed here, 39 were shifted to the available experimental values. The 5p5d $^3$F$^{\rm{o}}_4$ level retained its theoretical value, and its lack of strong transitions makes it unlikely to heavily impact any subsequent modelling.

\begin{figure}
    \centering
    \includegraphics[width=  \linewidth]{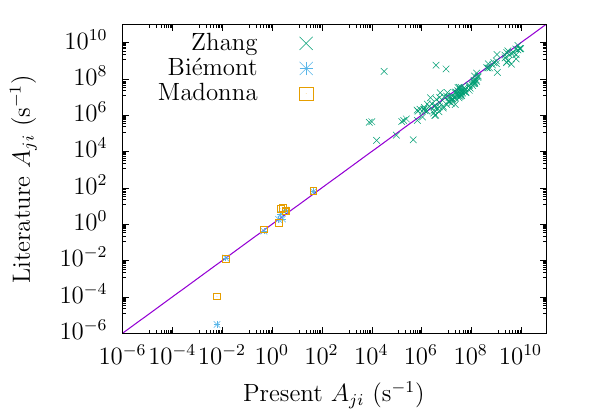}
    \caption{Present calculated values of $A_{ji}$ for Te {\sc iii} as compared with the the theoretical calculations of \citet{zhang2013transition} (E1 transitions), \citet{biemont1995forbidden} and \citet{madonna2018neutron} (E2 and M1 transitions).}
    \label{fig:avaluete2}
\end{figure}

No lines are listed in the NIST ASD for Te {\sc iii}. Potentially important forbidden transition probabilities are compared with the theoretical calculations of \citet{madonna2018neutron} and \citet{biemont1995forbidden} in Figure \ref{fig:avaluete2}.  With the exception of the seemingly problematic $^1$D$_2$ - $^3$P$_0$ transition, reasonable agreement is seen across the transitions listed. Additionally, strong dipole transition strengths were published by \citet{zhang2013transition}, which we also compare to in Figure \ref{fig:avaluete2}. It is clear that these E1 transitions agree very well between the data sets with few outliers. 
\begin{table}
    \centering
    \begin{tabularx}{0.475\textwidth}{lllllrr}
        \toprule
        \hfil Index & \hfil CSF & \hfil Level & \hfil Expt   & {\sc grasp$^0$} & $\Delta E$\hphantom{-} & \%\hphantom{-} \\ 
        \midrule
        1 & 5s$^2$ 5p$^2$            & $^3$P$_0$ & 0.0000  & 0.0000& 0.0000 & 0.00 \\ 
        2 & 5s$^2$ 5p$^2$            & $^3$P$_1$ & 0.0434  & 0.0378& -0.0055 & -12.78 \\ 
        3 & 5s$^2$ 5p$^2$            & $^3$P$_2$ & 0.0744  & 0.0718& -0.0026 & -3.47 \\ 
        4 & 5s$^2$ 5p$^2$            & $^1$D$_2$ & 0.1582  & 0.1701& 0.0119 & 7.54 \\ 
        5 & 5s$^2$ 5p$^2$            & $^1$S$_0$ & 0.2770  & 0.2961& 0.0191 & 6.91 \\ 
        6 & 5s \hspace{1.4mm}5p$^3$  & $^5$S$^{\rm{o}}_2$ & 0.5886  & 0.5322& -0.0564 & -9.58 \\ 
        7 & 5s \hspace{1.4mm}5p$^3$  & $^3$D$^{\rm{o}}_1$ & 0.7553  & 0.7471& -0.0082 & -1.09 \\ 
        8 & 5s \hspace{1.4mm}5p$^3$  & $^3$D$^{\rm{o}}_2$ & 0.7582  & 0.7495& -0.0087 & -1.15 \\ 
        9 & 5s \hspace{1.4mm}5p$^3$  & $^3$D$^{\rm{o}}_3$ & 0.7765  & 0.7663& -0.0101 & -1.31 \\ 
        10 & 5s$^2$ 5p 5d            & $^1$D$^{\rm{o}}_2$ & 0.8660   & 0.8678 &0.0019 & 0.21 \\ 
        11 & 5s \hspace{1.4mm}5p$^3$ & $^3$P$^{\rm{o}}_0$ & 0.8754   & 0.8834 &0.0081 & 0.92 \\ 
        12 & 5s \hspace{1.4mm}5p$^3$ & $^3$P$^{\rm{o}}_1$ & 0.8801   & 0.8860 &0.0059 & 0.67 \\ 
        13 & 5s \hspace{1.4mm}5p$^3$ & $^3$P$^{\rm{o}}_2$ & 0.9155   & 0.9170 &0.0015 & 0.16 \\ 
        14 & 5s$^2$ 5p 5d            & $^3$F$^{\rm{o}}_2$ & 0.9543   & 0.9692 &0.0150 & 1.57 \\ 
        15 & 5s$^2$ 5p 5d            & $^3$F$^{\rm{o}}_3$ & 0.9688   & 0.9856 &0.0168 & 1.74 \\ 
        16 & 5s$^2$ 5p 6s            & $^3$P$^{\rm{o}}_0$ & 0.9793   & 1.0342 &0.0549 & 5.60 \\ 
        17 & 5s$^2$ 5p 6s            & $^3$P$^{\rm{o}}_1$ & 0.9817   & 1.0379 &0.0563 & 5.73 \\ 
        18 & 5s$^2$ 5p 5d            & $^3$F$^{\rm{o}}_4$ & \hfil -  & 1.0257 &\hphantom{11}-&  \hphantom{-0..}-\hphantom{11}\\ 
        19 & 5s \hspace{1.4mm}5p$^3$ & $^3$S$^{\rm{o}}_1$ & 1.0408   & 1.1811 &0.1402 & 13.47 \\ 
        20 & 5s$^2$ 5p 6s            & $^3$P$^{\rm{o}}_2$ & 1.0518   & 1.1113 &0.0595 & 5.65 \\ 
        21 & 5s$^2$ 5p 6s            & $^1$P$^{\rm{o}}_1$ & 1.0548   & 1.1055 &0.0507 & 4.81 \\ 
        22 & 5s$^2$ 5p 5d            & $^3$D$^{\rm{o}}_2$ & 1.0636   & 1.1414 &0.0778 & 7.31 \\ 
        23 & 5s$^2$ 5p 5d            & $^3$D$^{\rm{o}}_1$ & 1.0734   & 1.1414 &0.0680 & 6.33 \\ 
        24 & 5s$^2$ 5p 5d            & $^3$D$^{\rm{o}}_3$ & 1.1018   & 1.1747 &0.0730 & 6.62 \\ 
        25 & 5s$^2$ 5p 5d            & $^3$P$^{\rm{o}}_1$ & 1.1129   & 1.1988 &0.0859 & 7.72 \\ 
        26 & 5s$^2$ 5p 5d            & $^3$D$^{\rm{o}}_2$ & 1.1164   & 1.2007 &0.0843 & 7.55 \\ 
        27 & 5s$^2$ 5p 5d            & $^3$P$^{\rm{o}}_0$ & 1.1167   & 1.1975 &0.0808 & 7.23 \\ 
        28 & 5s$^2$ 5p 5d            & $^3$P$^{\rm{o}}_1$ & 1.1372   & 1.2306 &0.0935 & 8.22 \\ 
        29 & 5s \hspace{1.4mm}5p$^3$ & $^1$D$^{\rm{o}}_2$ & 1.1590   & 1.2772 &0.1182 & 10.20 \\ 
        30 & 5s$^2$ 5p 5d            & $^1$F$^{\rm{o}}_3$ & 1.1595   & 1.2285 &0.0690 & 5.95 \\ 
        31 & 5s$^2$ 5p 6p            & $^3$D$_1$ & 1.1721   & 1.1711 &-0.0010 & -0.08 \\ 
        32 & 5s$^2$ 5p 6p            & $^3$P$_1$ & 1.2039   & 1.2058 &0.0019 & 0.16 \\ 
        33 & 5s$^2$ 5p 6p            & $^3$P$_0$ & 1.2053   & 1.2193 &0.0141 & 1.17 \\ 
        34 & 5s$^2$ 5p 6p            & $^3$D$_2$ & 1.2059   & 1.2116 &0.0057 & 0.48 \\ 
        35 & 5s \hspace{1.4mm}5p$^3$ & $^1$P$_1^{\rm{o}}$ & 1.2437   & 1.3452 &0.1016 & 8.17 \\ 
        36 & 5s$^2$ 5p 6p            & $^3$P$_1$ & 1.2602   & 1.2581 &-0.0021 & -0.17 \\ 
        37 & 5s$^2$ 5p 6p            & $^3$P$_2$ & 1.2727   & 1.2783 &0.0056 & 0.44 \\ 
        38 & 5s$^2$ 5p 6p            & $^3$D$_3$ & 1.2753   & 1.2721 &-0.0032 & -0.25 \\ 
        39 & 5s$^2$ 5p 6p            & $^3$S$_1$ & 1.2922   & 1.2930 &0.0008 & 0.06 \\ 
        40 & 5s$^2$ 5p 6p            & $^1$D$_2$ & 1.3029   & 1.3350 &0.0321 & 2.46 \\ 
        \bottomrule
    \end{tabularx}
    \caption{Te {\sc iii} energy levels in Rydbergs, as compared with the experimental results of \citet{tauheed2011revision,joshi1992analysis} and identifications by \citet{zhang2013transition}}
    \label{table:energy_levels_double}
\end{table}

\section{Electron-Impact Excitation}\label{sec:electron_impact_excitation}

A comprehensive overview of the details surrounding Dirac-R-matrix theory is given in \citet{burke2011r}. Here, we employ the {\sc darc} \citep{Ballance2020_DARC} package. This package is interfaced with {\sc grasp$^0$} in our workflow. For a target with $N$ electrons, the numerical treatment of electron-impact-excitation involves representing and diagonalizing the ($N+1$) Hamiltonian to extract the dimensionless collision strengths $\Omega_{ij}$ for a transition $i\to j$. To this end, the close-coupling  expansion is used to represent the ($N+1$) system. For modelling purposes, such quantities are often thermally averaged according to some energy distribution. In this work we consider  Maxwellian-averaged collision strengths  given by,
\begin{equation}
    \Upsilon_{ij}(T_e) = \int_0^{\infty} \Omega_{ij}(\epsilon_j)e^{-\epsilon_j / kT_e}\hspace{1mm} \text{d}\left(\epsilon_j / kT_e\right) \label{eq:effective_cs},
\end{equation}
where $T_e$ is the temperature of the electron gas, $k$ is the Boltzmann constant and $\epsilon_j$ is the electron energy after impact. This is often also referred to as an effective collision strength. To model the considered plasma, one then calculates the excitation $q_{i\to j}$ and de-excitation $q_{j\to i}$ collision rates  via
\begin{align*}
       q_{i\to j}(T_e) &= \frac{8.63\times10^{-6}}{g_i T_e^{1/2}} \Upsilon_{ij}(T_e) e^{-E_{ij}/kT_e} \hspace{1mm}\label{eq:rates}, \\
    q_{j \to i}(T_e) &= \frac{g_i}{g_j} e^{E_{ij} /kT_e} q_{i\to j},
\end{align*}
 in units of $\text{cm}^3\hphantom{1}\text{s}^{-1}$. Here $g_i$ is the statistical weight of the lower level and $E_{ij}$ is the energy difference between the two states.

In such calculations the collision strength is given by the sum of the contributions from each $J\pi$ symmetry of the ($N+1$) system (referred to as partial waves), with $J$ the total angular momentum and $\pi$ the parity. Such a sum is naturally truncated for practical implementation, but to approximate the contribution from high partial waves we employ the `top-up' procedure of \citet{burgess1974coulomb}. To calculate the integral in Equation~\eqref{eq:effective_cs}, the collision strengths for electric-dipole transitions are extrapolated to an infinite energy point using the limiting behaviours outlined by \citet{burgess1992analysis}. Additionally, the diagonal elements of the Hamiltonian are shifted to spectroscopic energies where possible to ensure the correct position of resonances in the cross sections. 

For the remainder of this section, we describe the most pertinent parameters in the electron impact excitation calculation for each considered ion and present both collision strength and effective collision strength profiles for some potentially interesting transitions. The calculated rates are then used in a collisional radiative model to be outlined in Section \ref{sec:colrad}. The calculated $\Upsilon_{ji}$ will be made available in the standard {\sc adf04} format at \cite{openadas_site}.

\subsection{Te {\sc i}}
\begin{figure*} 
\centering
    \includegraphics[width = 0.8 \linewidth]{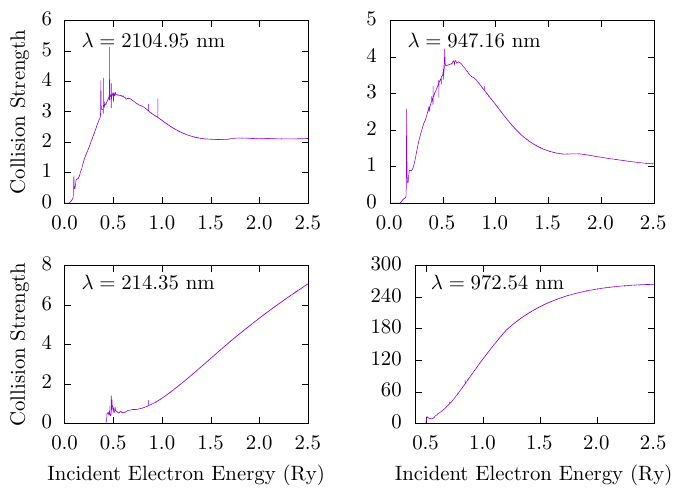}
    \caption{Collision strengths for four Te {\sc i} transitions. The transitions presented are  5s$^2$ 5p$^4$  $^3$P$_1$ $\to$ 5s$^2$ 5p$^4$  $^3$P$_2$ ($\lambda = 2104.95$ nm); 5s$^2$ 5p$^4$  $^1$D$_2$ $\to$ 5s$^2$ 5p$^4$  $^3$P$_2$ ($\lambda = 947.16$ nm);  5s$^2$ 5p$^3$ 6s $^3$S$_1^\mathrm{o}$ $\to$ 5s$^2$ 5p$^4$  $^3$P$_2$ ($\lambda = 214.35$ nm) and  5p$^3$ 6p $^5$P$_3$ $\to$ 5p$^3$ 6s $^5$S$_2^\mathrm{o}$ ($\lambda = 972.54$ nm).}
    \label{fig:Te0+:background-cs}
\end{figure*}
\begin{figure*}
    \centering
    \includegraphics[width = 0.8 \linewidth]{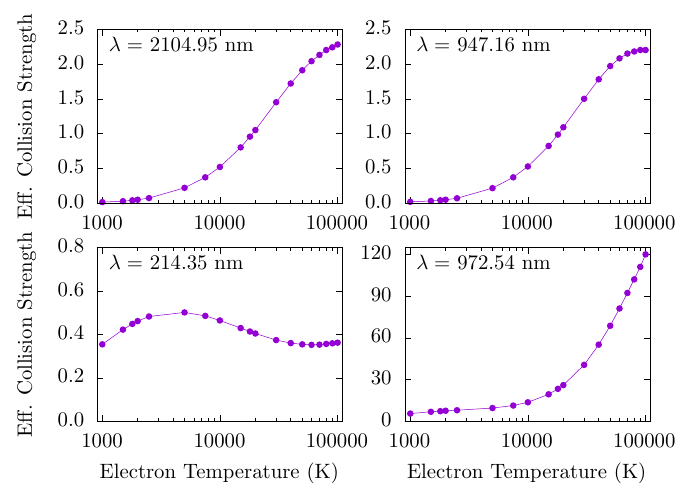}
    \caption{Collision strengths for four Te {\sc i} transitions. The transitions presented are the same as those in Figure \ref{fig:Te0+:background-cs}.}
    \label{fig:Te0+:maxwell-cs}
\end{figure*}

For neutral tellurium, 150 fine structure levels were retained in the close-coupling collision calculations. The R-matrix boundary was set at $r = 44.16$ atomic units, ultimately requiring diagonalization of matrices of size up to $31429 \times 31429$. Given the particularly strong low lying transitions, we found that a total of 80 $J\pi$ partial waves was required for convergence. We employed a continuum orbital basis size of 45 for $2J = 1 - 80$. This was sufficient to span the relatively large energy range of 0 - 5 Ry. While such high energies are unlikely to be of importance under KNe conditions, a complete structure of the collision profile provides both convergence of the collision rates and the possibility for the data to be used to benchmark higher energy phenomena. The resonance structure was captured by using a fine mesh of 22800 points for $2J = 1-9$ and 11400 points for $2J = 11-29$ with an energy spacing of $2.19\times 10^{-4}$ Ry and $4.88\times 10^{-4}$ Ry respectively. For the higher partial waves a coarse mesh of 1024 points with an energy spacing of $4.88\times 10^{-3}$ Ry was used. 

In Figure \ref{fig:Te0+:background-cs} we present the collision-strengths $\Omega$ for four transitions of interest. In particular, we show the diagonostically important \citep{gillanders2024modelling,hotokezaka2023tellurium} 5p$^4$ $^3$P$_1$ $\to$ $^3$P$_2$  ($2104.95$ nm) line. We also show: 5p$^4$ $^1$D$_2$  $\to$ $^3$P$_2$ (947.16 nm); 5p$^3$6s $^3$S$_1^{\rm{o}}$  $\to$ 5p$^4$ $^3$P$_2$ (214.35 nm) and 5p$^3$6p $^5$P$_3$  $\to$ 5p$^3$6s $^5$S$_2^{\rm{o}}$ (972.54 nm). Three of these are particularly strong low lying transitions, with two being forbidden. 

In Figure \ref{fig:Te0+:maxwell-cs} we present the thermally averaged effective collision strengths for the same four transitions, which will be employed in Section \ref{sec:colrad} in a collisional radiative model.

\subsection{Te {\sc ii}}
\begin{figure*}
    \centering
    \includegraphics[width = 0.8 \linewidth]{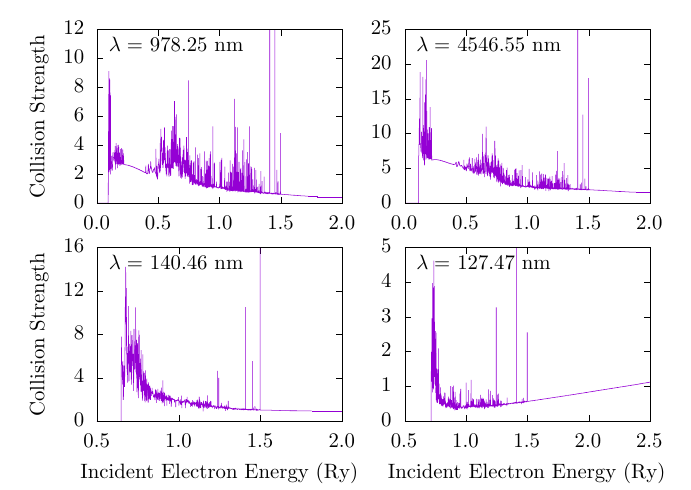}
    \caption{Collision strengths as a function of incident electron energy for four Te {\sc ii} transitions. The presented transitions are from the transitions: 5s$^2$5p$^3$ $^4$S$^{\mathrm{o}}_{3/2}$ $\to$ 5s$^2$5p$^3$ $^2$D$^{\mathrm{o}}_{3/2}$  (978.25 nm); 5s$^2$5p$^3$ $^2$D$^{\mathrm{o}}_{3/2}$ $\to$ 5s$^2$5p$^3$ $^2$D$^{\mathrm{o}}_{5/2}$  (4546.55 nm) ; 5s$^2$5p$^3$ $^4$S$^{\mathrm{o}}_{3/2}$ $\to$ 5s5p$^4$ $^4$P$_{5/2}$  (140.46 nm) and 5s$^2$5p$^3$ $^4$S$^{\mathrm{o}}_{3/2}$ $\to$ 5s$^2$5p$^2$6s $^4$P$_{1/2}$  (127.47 nm).}
    \label{fig:Te1+:background-cs}
\end{figure*}

\begin{figure*}
    \centering
    \includegraphics[width = 0.8 \linewidth]{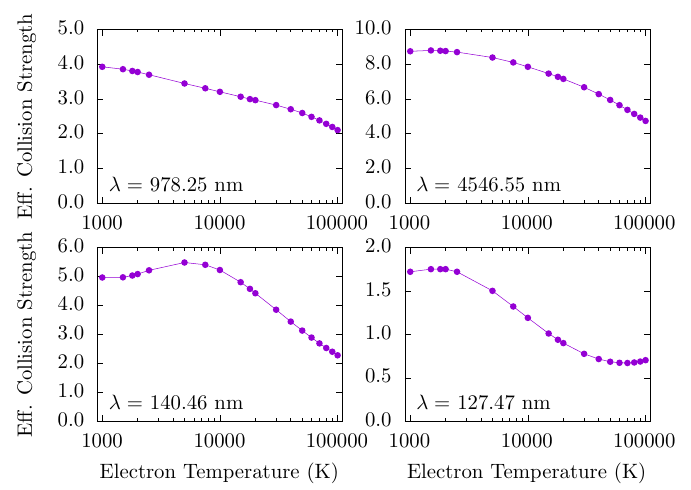}
    \caption{Maxwellian averaged effective collision strengths for Te {\sc ii}. The transitions presented are the same as those in Figure \ref{fig:Te1+:background-cs}.}
    \label{fig:Te1+:maxwell-cs}
\end{figure*}

In these collision calculations we included 150 fine structure levels in the close coupling expansion of the target wavefunction. A continuum orbital basis of size 25 and an R-matrix boundary of $29.58$ atomic units produced Hamiltonian matrices of size up to $17800 \times 17800$. A total of 64 $J\pi$ partial waves were explicitly included. For the lower partial waves with $2J = 0-30$, the resonance structure was captured using 25600 points on a  fine mesh energy spacing of $1.56\times 10^{-4}$ Ry. A coarse mesh of 2560 points with a spacing of $1.56\times10^{-3}$ Ry was used for the remaining partial waves, along with the top-up partial wave procedure \citep{burgess1974coulomb}. Figure \ref{fig:Te1+:background-cs} shows the collision strengths for the transitions 5s$^2$5p$^3$ $^4$S$^{\mathrm{o}}_{3/2}$ $\to$ 5s$^2$5p$^3$ $^2$D$^{\mathrm{o}}_{3/2}$; 5s$^2$5p$^3$ $^2$D$^{\mathrm{o}}_{3/2}$ $\to$ 5s$^2$5p$^3$ $^2$D$^{\mathrm{o}}_{5/2}$ ; 5s$^2$5p$^3$ $^4$S$^{\mathrm{o}}_{3/2}$ $\to$ 5s5p$^4$ $^4$P$_{5/2}$  and 5s$^2$5p$^3$ $^4$S$^{\mathrm{o}}_{3/2}$ $\to$ 5s$^2$5p$^2$6s $^4$P$_{1/2}$. The corresponding effective collision strengths are shown in Figure \ref{fig:Te1+:maxwell-cs}. Similarly for Te {\sc i}, there are no available data sets to compare with. In contrast to the neutral Te {\sc i}, the resonance structure in the calculated collision profiles is considerably more pronounced.

\subsection{Te {\sc iii}}
\begin{figure*}
    \centering
    \includegraphics[width = 0.8 \linewidth]{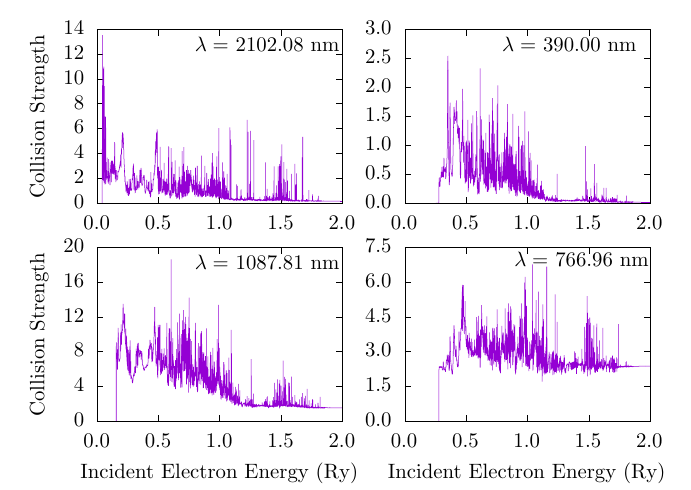}
    \caption{Collision strengths as a function of incident electron energy for four Te {\sc iii} transitions. The presented transitions are from the forbidden ground state transitions: $^3$P$_1$ $\to$ $^3$P$_0$  (2102.08 nm); $^1$S$_0$ $\to$ $^3$P$_1$ (390.00 nm); $^1$D$_2$ $\to$ $^3$P$_2$ (1087.81 nm) and $^1$S$_0$ $\to$ $^1$D$_2$ (767.96 nm).}
    \label{fig:te2+:background-cs}
\end{figure*}

\begin{figure*}
    \centering
    \includegraphics[width = 0.8 \linewidth]{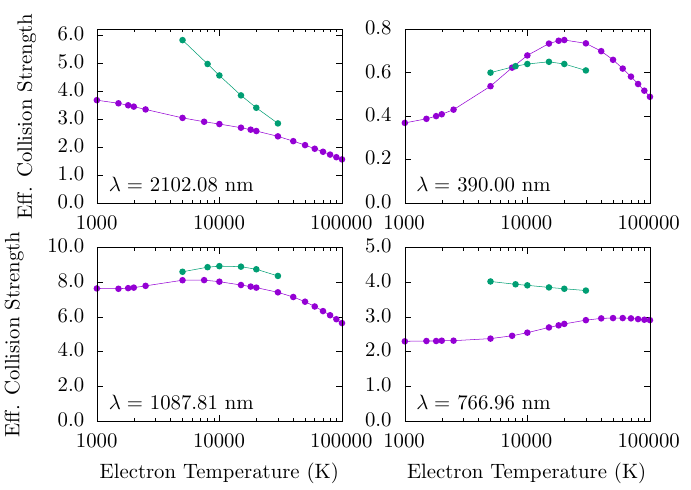}
    \caption{Effective collision strengths as a function of the electron temperature compared with the calculations of \citet{madonna2018neutron}. The presented transitions are the same as those presented in Figure \ref{fig:te2+:background-cs}.}
    \label{fig:te2+:maxwell-cs-madonna}
\end{figure*}
For the doubly-ionised case, a total of 100 levels were included in the close coupling expansion with an orbital basis size of 20. The R-matrix boundary is set at 19.84 atomic units. The Hamiltonian matrices had a maximum size of $9400 \times 9400$. There were 64 $J\pi$ waves included in the expansion (plus top-up). For $2J = 1 - 31$, a very fine mesh was used to ensure the resonance structure was adequately captured. In the energy interval [0.00,0.36] Ry, 9600 points were included with an energy spacing of $3.73\times10^{-5}$ Ry. In the higher energy interval of [0.36,2.70] Ry, 19200 points were used with an energy spacing of $1.35\times10^{-4}$ Ry. The remaining partial waves (plus top-up) had 1280 energy points with a spacing of $7.81\times10^{-4}$ Ry.

For this ion, data have been published by \citet{madonna2018neutron} and  used to identify Te {\sc iii} in planetary nebulae. That previous calculation was restricted to four CSFs and presents effective collision strengths for transitions within the ground state. In their study the semi-relativistic Breit-Pauli suite of codes were employed \citep{berrington1995rmatrx1}. In general, we find notable disagreement with their calculation, where in some cases the effective collision strengths differ by a factor of two or more. A comparison of four ground state transitions is shown on Figure \ref{fig:te2+:maxwell-cs-madonna}. It is clear that there is quite considerable disagreement between the two calculations, the source of error is unclear at this time.  Of particular note is the Te {\sc iii}  line at 2.1$\mu$m, where there is notable disagreement particularly at low temperatures, where this is up to a factor of 1.5 difference. As will be discussed in Section 5, mass estimates of Te {\sc iii} that have been made previously using low temperature predictions of this emission feature \cite{madonna2018neutron,hotokezaka2023tellurium,levan2024heavy} may be uncertain because of this.
Using a Breit-Pauli set of codes, and using the descriptions in \citet{madonna2018neutron} we were unable to reproduce their results.

\section{1D LTE {\sc TARDIS} modelling}\label{sec:tardis}

1D LTE spectral synthesis codes such as \textsc{tardis} \citep{KerzendorfSimTardis, kerzendorf_2023_8244935} have been recently used by \cite{gillanders2022}, \cite{sneppen2023discovery}, \cite{vieira2023}, and \cite{tak2024}, amongst others, to model the spectral evolution of the early epochs of AT2017gfo and to propose identifications for spectral features.  Here, we use \textsc{tardis} to visualise the effect our new atomic calculation has on synthetic spectra with a comparison to previously used data sets.

Following our previous analysis using newly calculated atomic data for Sr {\sc ii} and Y {\sc ii} \citep{Mulholland24} and using the same parameters as found in table 9 of their work, we again replicate the methods of \cite[][hereafter G22]{gillanders2022} using \textsc{tardis} to generate synthetic spectra to match early time observations of AT2017gfo, this time using our new calculated atomic data for Te. We again focus on the 1.4 and 4.4 days post-merger epochs, as these are still well approximated by a blackbody continuum, as assumed by \textsc{tardis}. We use two datasets: one constructed from the same sources as G22, and one where we replace the previously used Te data with that presented in this paper.

\begin{figure*}
    \centering
    \includegraphics[width = \linewidth]{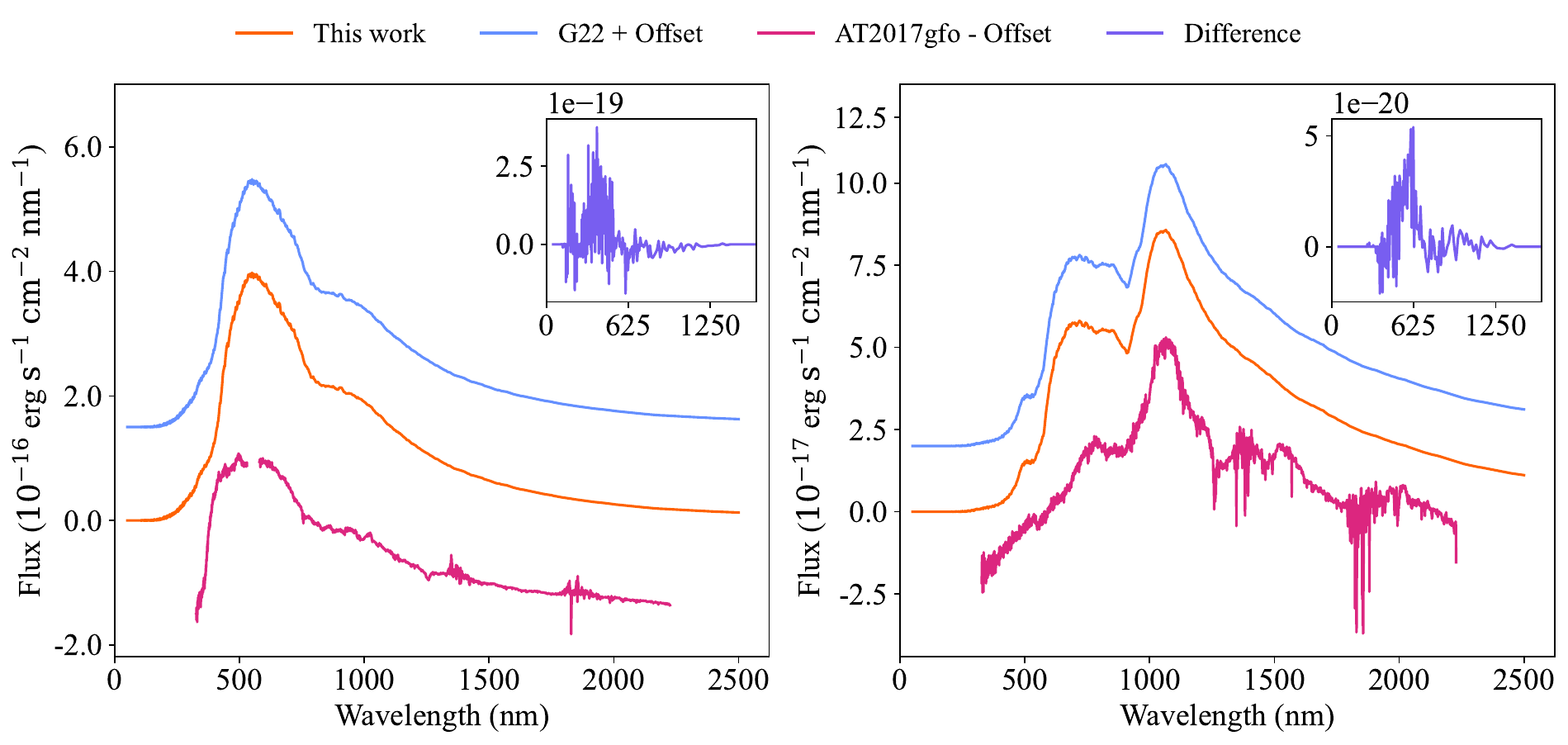}
    \caption{\textsc{tardis} models of AT2017gfo at 1.4 (left) and 4.4 (right) days post-merger. Our synthetic spectra are presented in comparison to observations of AT2017gfo published by \citet{Pian17} and \citet{smartt2017kilonova}, with inset plots of the difference between using the G22 dataset and the new calculation from this work. The observations and synthetic spectra from the G22 dataset have been arbitrarily offset from the synthetic spectra constructed using this new calculation for Te, by $\mp$ 1.5 E-16 erg s$^{-1}$ cm$^{-2}$ nm$^{-1}$ at 1.4 days and $\mp$ 2.0 E-17 erg s$^{-1}$ cm$^{-2}$ nm$^{-1}$ at 4.4 days. We note that there is a visual difference between our synthetic spectra and the observations, due to the updated relativistic treatment in \textsc{tardis} \citep{vogl2019}. Both datasets are treated using this updated relativity to allow comparison, as described in \citet{Mulholland24}. }
    \label{fig:tardis_line_plots}
\end{figure*}

It can be seen from \autoref{fig:tardis_line_plots} that there is very little difference seen in the synthetic spectra between the datasets: the two are nearly identical within the Monte Carlo noise inherent to such simulations. The noise levels in the difference plots tell a similar story, subtracting the two datasets from each other results in mostly random noise. This is not unexpected: the 2.1 $\mu$m Te line predicted by \cite{hotokezaka2023tellurium} is prominent beyond the time frame in which the photospheric approximations used by \textsc{tardis} remain valid, and there is little proposed Te contribution seen in the observations in the early phases modelled here. It is possible that with a future push towards developing NLTE models of AT2017gfo that a bigger contribution from Te may present itself at later times: such as the 7.5-10.5 day spectra discussed by \cite{hotokezaka2023tellurium}, but this regime requires collisional modelling codes, as discussed in \autoref{sec:colrad} below.

\section{NLTE Collisional-Radiatve modelling} \label{sec:colrad}

As a first use of the newly acquired collision data, we employ a collisional radiative model \citep{bates1962recombination} implemented in the ColRadPy package \citep{johnson2019colradpy}. For each ion, the level populations $N_i$ are calculated according to the set of differential equations,
\begin{equation}
    \frac{\dd N_i}{\dd t} = \sum_{j} C_{ij} N_j,\label{eq:cr}
\end{equation}
where $C_{ij}$ is a collisional-radiative matrix encompassing the pertinent excitation and de-excitation rates.  We are thus able to produce theoretical (optically thin in this case) spectra, emission line strengths and ratios, and level populations as a function of electron temperature and density. We present these results in the remainder of this section.

\subsection{Level Populations}

\begin{figure*}
    \centering
    \includegraphics[width =  \linewidth]{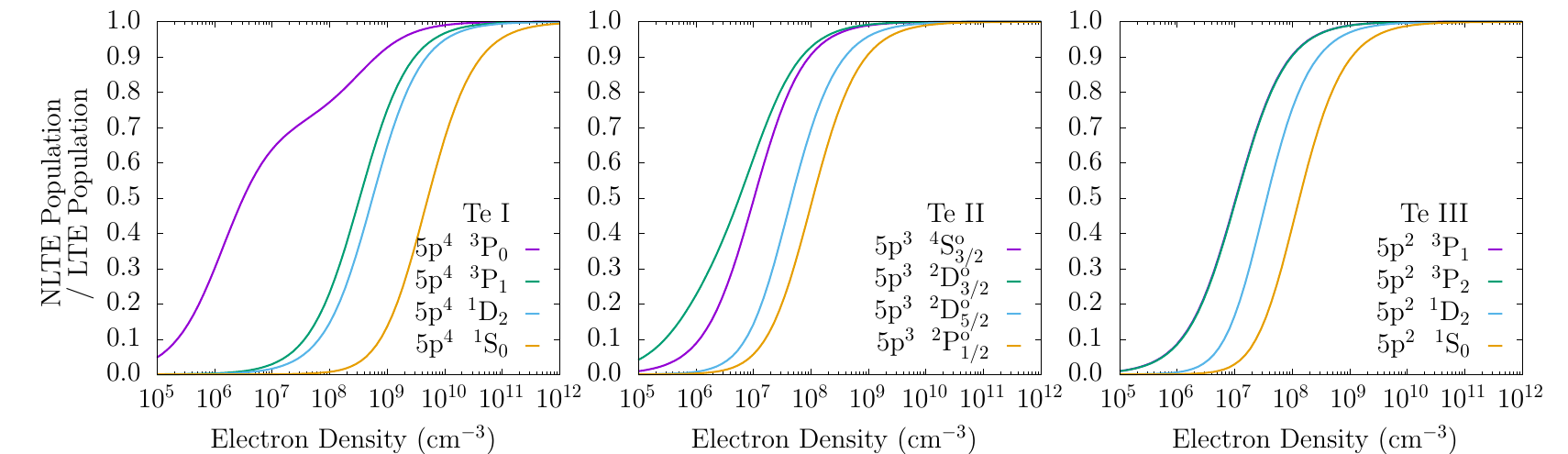}
    \caption{NLTE population fractions divided by the corresponding LTE poulation fraction for the first few excited states of Te {\sc i - iii} as a function of electron density. The electron temperature is set to 0.2 eV.}
    \label{fig:te_all_three_pops}
\end{figure*}

It is common practice in modelling to assume the approximation of Local-Thermodynamic-Equilibrum (LTE). In this regime, the populations of the atomic levels $N_i$ follow a simple Boltzmann distribution with,
\begin{equation}
    \frac{N_j}{N_i} = \frac{g_j}{g_i} e^{ - (E_j - E_i) /{kT_e}},
\end{equation}
where $i$ and $j$ are level indices, $g_{i,j}$ are the statistical weights of the levels, and $E_{i,j}$ their energies. This circumvents the need for any collision excitation rates as the populations are entirely determined by temperature. This is a valid approximation at high electron densities, where the collision rates outweigh the Einstein A-coefficients of spontaneous emission. LTE can also be reached in early phase KNe modelling where radiation is dominant.  While LTE in general leads to more accessible computational implementations, it is expected to be inadequate at the particularly low temperatures and densities present in the late stage nebular phases of KNe. Using the ColRadPy \citep{johnson2019colradpy} implementation, we are able employ the previously calculated rates to calculate populations using a collisional radiative model \citep{bates1962recombination,Summers2006}.

We show the population behaviour of the first four excited states of Te {\sc i}-{\sc iii} as a function of electron density in Figure \ref{fig:te_all_three_pops} at a temperature of $0.2$ eV. Here the populations are divided by the LTE populations to show the deviation from LTE.  Evidently, LTE is reached for these low lying levels of the three ions at relatively high densities above $10^9$ cm$^{-3}$. By contrast there is significant deviation from LTE at the densities of concern for KNe work ($10^6 - 10^7$ cm$^{-3}$), where the populations are in general overestimated by LTE. This in principle could lead to inaccurate emission from higher levels in modelling codes. Furthermore, it is interesting that the first excited state of Te {\sc i}, $5$p$^4$ $^3$P$_0$ exhibits considerably different behaviour than the other states shown, and reaches LTE much earlier in electron-density space. This is likely due to the fact that this level has a weak decay to the ground state with transition probability $\sim 1.53 \times 10^{-2}$ s$^{-1}$. The contribution from radiative decays is therefore easily outweighed by the collision rates at low densities. By contrast, the other featured states have relatively stronger decays to the ground with larger $A$-values.

\subsection{Synthetic Emission Line Spectra}

\begin{figure*}
    \centering
    \includegraphics[width = \linewidth]{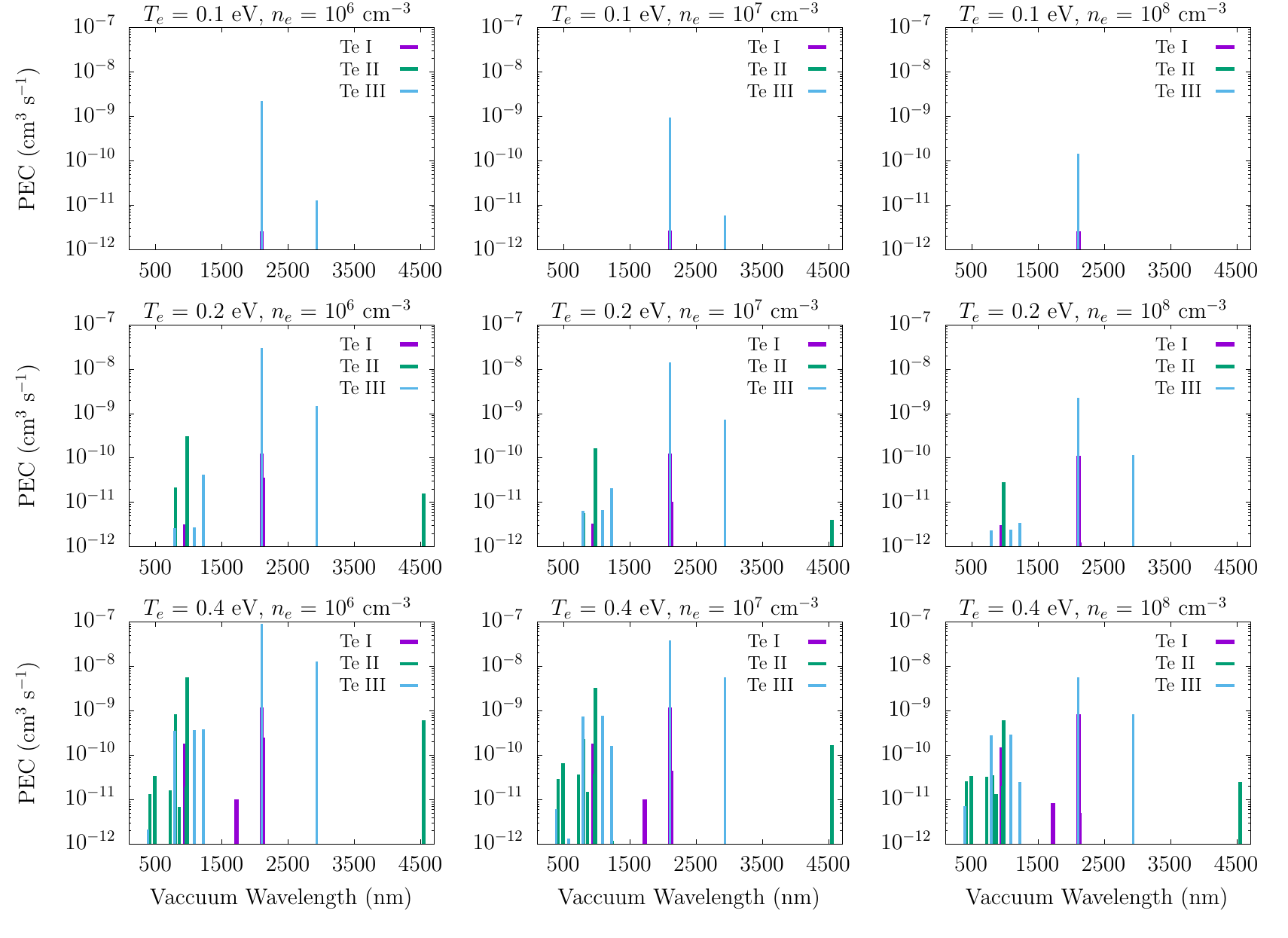}
    \caption{Synthetic excitation spectra for Te {\sc i}-{\sc iii} at $T_e \in \{0.1,0.2,0.4\}$ eV and electron density $n_e \in \{10^6,10^7,10^8\}\phantom{1}\text{cm}^{-3}$.}
\label{fig:colradpy:full_spec_many_conditions_landscape}
\end{figure*}

Using the calculated populations, we are able to produce synthetic optically thin emission spectra for each of the three ion stages. These are most useful for interpreting late time observation of KNe which are collisionally dominated. These spectra have been calculated in terms of photon-emissivity-coefficients (PEC) given by,
\begin{equation}
    \text{PEC}_{i\to j} = \frac{N_i A_{i\to j}}{n_e},
\end{equation}
where $N_i$ is taken relative to the ground. These can then be used to calculate photon luminosities via,

\begin{equation}
    L = n_e  \frac{\text{PEC}_{i\to j}}{\sum_i N_i}  \frac{M_{\text{ion}}}{\text{m}_{\text{ion}}} \frac{hc}{\lambda}, \label{eq:photon_lumo}
\end{equation}
in units of erg s$^{-1}$. Here $M_{\text{ion}}$ is the total mass of the ion in the ejecta and $\text{m}_{\text{ion}}$ is the mass of a single ion particle. Here $h$ and $c$ are the Planck constant and speed of light respectively, and $\lambda$ is the transition wavelength.  We have explored the parameter space $n_e \in \{10^6,10^7,10^8\}\phantom{1}\text{cm}^{-3}$ and temperatures $T_e \in \{0.1,0.2,0.4\}$ eV. The PECs are shown in Figure \ref{fig:colradpy:full_spec_many_conditions_landscape}. 

At these conditions, that are expected to be approximately indicative of those in the nebular phases of the KNe, we predict the strongest line to be from Te {\sc iii} at 2.1 $\mu$m, in agreement with the predictions of \citet{hotokezaka2023tellurium,gillanders2024modelling,gillanders2023heavy}. At high temperatures and densities, there is notable overlap at approximately the same wavelength from Te {\sc i}, however, the Te {\sc iii} line is predicted to be upwards of an order of magnitude stronger than that of Te {\sc i}.  While the Te {\sc ii} line at 4.4 $\mu$m is recovered, it has a comparatively weak PEC.   Interestingly, contributions from Te {\sc ii} generally only become noticeably strong when compared with those from Te {\sc iii} at temperatures above 0.2 eV $\sim$ 2300 K. The dominant contributions from Te {\sc ii} appear to be in the visible interval of the spectrum, in contrast to the dominant near-infrared contributions from Te {\sc i} and {\sc iii}.  In general, additional features begin to appear with increasing temperature as more levels become thermally accessible for excitation and emission.

To demonstrate the relative strengths of the lines, luminosities have been calculated for a reference mass of  $10^{-3} M_\odot$ for each ion stage, assigned to each ion in Table \ref{tab:lumo} at two conditions, typical of the AT2017gfo at 10.4 days and GRB230307A at 29 days respectively. The ten strongest lines from each ion are listed. 

For Te {\sc iii} to explain the feature at $\sim$ 2.1$\mu$m line in the AT2017gfo, a rough mass of $\sim 10^{-3} M_\odot $ of this ion has been calculated by \cite{hotokezaka2023tellurium} using $T_e = 0.17$ eV and LTE to produce a total luminosity of $\sim 2 \times 10^{39}$ erg s$^{-1}$. Their calculation was supplemented by an optically thin NLTE model, including electron-impact-excitation/de-excitation and spontaneous emission and ionization balance \citep{Hotokezaka2022WSe,hotokezaka2023tellurium}, with a large multi-species linelist using $T_e = 0.17$ eV and $n_e = 7.4 \times 10^{6}$ cm$^{-3}$ at 10.4 days. Employing these same extrinsic parameters requires a mass of $2.8 \times 10^{-3} M_\odot $, using our  new atomic data and the assumptions of optically thin emission and excitation dominated by collisions, to produce the same luminosity value using Eq. \eqref{eq:photon_lumo}. By contrast, the atomic data of \cite{madonna2018neutron} requires a mass of $2.0 \times 10^{-3} M_\odot $ in using our model. The larger mass requirement of the current atomic data is consistent with the difference in the effective collision strengths highlighted in Figure \ref{fig:te2+:maxwell-cs-madonna}. In either case, it is clear that an LTE assumption underestimates the mass by overestimating the level populations, and given the ad-hoc approximations made here, the three calculations are in relatively good agreement. 

Using the absolute luminosities calculated here and presented in Table \ref{tab:lumo}, one can develop specific luminosities by constructing Guassian line profiles whose integral gives the specified absolute luminosity. The Gaussian broadening parameter was set to $0.07c$. One can then compare directly to the observed spectra. The mass of Te {\sc iii} was scaled to the previously discussed value of $2.8 \times 10^{-3} M_\odot $ while the masses of the other ions were scaled to follow the mass ratio of (0.25,0.4,0.25,0.1) in accordance with \cite{hotokezaka2023tellurium}. In reality ionization balance is itself a time-dependent phenomenon (see e.g calculations of \citet{pognan_steady}) that should be taken into account in Eq. \eqref{eq:cr} as data for ionization and recombination becomes available, and could in principle have an effect on the relative strengths of the Te features in synthetic emission. 

\begin{figure}
    \centering
    \includegraphics[width = \linewidth]{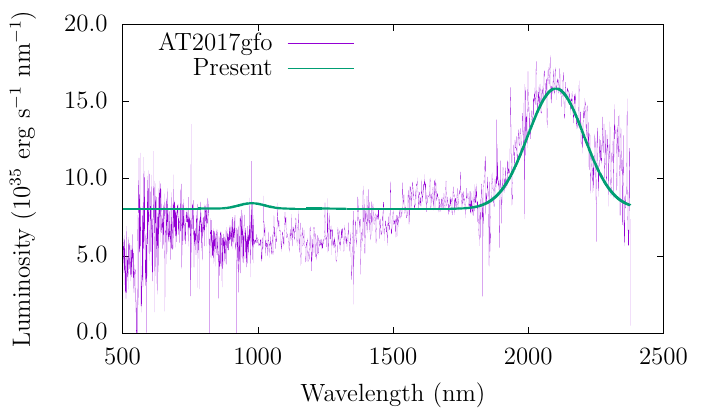}
    \caption{
    Spectrum of AT2017gfo at +10.4 days, 
    compared to the present Te spectrum, where the lines calculated (the strongest of which are in  Table \ref{tab:lumo}) has been given a Gaussian broadening parameter of $0.07c$.}
    \label{fig:lumo_obs}
\end{figure}

\begin{figure}
    \centering
    \includegraphics[width = \linewidth]{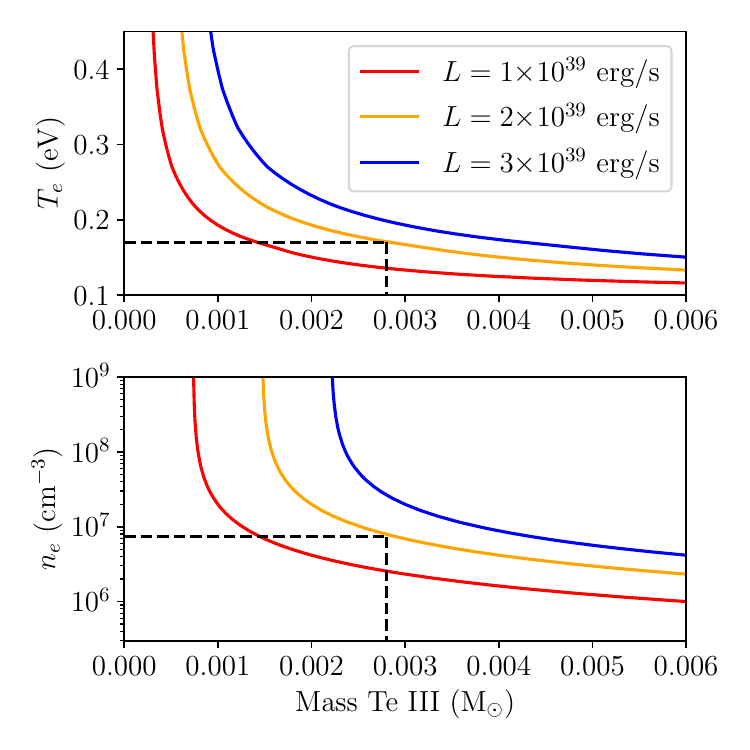}
    \caption{Contour lines of constant luminosity of the 2.1$\mu$m Te {\sc iii} line. On the top panel, the electron density is fixed at $n_e = 7.4 \times 10^{6}$ cm$^{-3}$ and the electron temperature and mass of Te {\sc iii} parameter spaces are explored. The dashed straight lines indicate $T_e = 0.17$eV and a mass of $2.8 \times 10^{-3} M_\odot $. On the bottom panel, the electron temperature is fixed at 0.17 eV and the electron density and mass of Te {\sc iii} parameter spaces are explored. The dashed straight lines indicate $n_e = 7.4 \times 10^{6}$ cm$^{-3}$ and a mass of $2.8 \times 10^{-3} M_\odot $.}
    \label{fig:contour}
\end{figure}

A model spectrum was produced for comparison with the +10.4d spectrum captured using the X-Shooter spectrograph at the European Southern Observatory's Very Large Telescope \citep{Pian17, smartt2017kilonova}, using the absolute luminosities presented in Table \ref{tab:lumo}. We focus on this relatively late epoch as the spectrum is plausibly optically thin, and collisionally dominated excitations can be safely assumed. It is of note that this feature can be discerned from earlier epochs as well where the validity of these assumptions is less pronounced \cite{sneppen2024emergencehourbyhourrprocessfeatures}.   This is shown on Figure \ref{fig:lumo_obs}. Assuming a continuum luminosity of $0.8\times 10^{36}$ erg s$^{-1}$ Å$^{-1}$ \citep{gillanders2024modelling}, the profile is seen to fit the observation with reasonable parameters. This seems to suggest that Te {\sc iii} could soley produce this feature, however the mass required is perhaps quite high and the profile relatively wide. We therefore cannot exclude that a blend is needed from this analysis alone, which has also been proposed in the literature by \citep{gillanders2024modelling} in terms of the blend of two Gaussian profiles which is motivated by the seemingly larger measured velocity than that of the 1.6 $\mu$m feature. Nevertheless, this basic analysis allows for a rough over-estimation of the mass of Te {\sc iii} by assuming a single profile, although does not rule out a double-line-profile that is also proposed in the literature. In addition to this analysis, we present on Figure \ref{fig:contour} contour lines of constant luminosity, as a function of the electron temperature and mass of Te {\sc iii} with the electron density fixed at $n_e = 7.4 \times 10^{6}$ cm$^{-3}$, and again at a fixed temperature of 0.17 eV and varying density. These were calculated using the present {\sc darc} atomic data and the modelling assumptions of Section 5.2. This reveals how the required mass and electron temperature/density can vary based on changes to the luminosity. The lower luminosity value is particularly important in this case, where the potential for a blend of two features in the observation has been proposed here would reduce the required mass of Te {\sc iii} and thus potentially redirect the estimates of electron temperature and mass in this case. The behaviour in density space at a fixed electron temperature is similar

The synthetic spectrum also predicts a small contribution from the forbidden line of Te {\sc ii} at  0.978 $\mu$m. Despite the potential overabundance of Te {\sc ii} in this model, this artefact is considerably weaker than the 2.1 $\mu$m feature. Note that the model lies considerably above the observation here on Figure \ref{fig:lumo_obs} since the model crudely approximates the continuum everywhere using the fitted value at the 2.1$\mu$m line from \cite{gillanders2024modelling}. While the X-shooter spectrum is interesting in this wavelength vicinity, the lack of a meaningful feature at this wavelength in the observation weakly constrains the mass of Te {\sc ii} in the KNe to be $\lessapprox 5\times 10^{-3} \text{M}_{\odot}$ assuming a similar electron temperature and density.

Of additional interest is the Spitzer observation at $4.5 \mu$m, reported by \cite{Villar_2018_spitzer,Kasliwal_spitzer}. These late time observations of the KNe report bolometric luminosities in the IRAC band of $\sim 6 \times 10^{38}$ erg s$^{-1}$ and $\sim 2 \times 10^{38}$ erg s$^{-1}$ at +43d and +74d respectively. It has been suggested by \cite{gillanders2024modelling} that the transition Te {\sc ii} $^2$D$^{\rm{o}}_{3/2}$ $\to$ $^2$D$^{\rm{o}}_{5/2}$ ($\lambda = 4.547 \mu$m) could contribute to a feature in this wavelength range. Based on the collision data published here, a significant contribution is unlikely. This line is predicted to be weak at the +10.4d conditions under the chosen conditions and to produce the total luminosity we would require an unphysical $\sim 5$ $\text{M}_{\odot}$ of Te {\sc ii}. We should expect cooler and sparser conditions by the late times reported in the Spitzer observations, making Te {\sc ii} an unconvincing candidate for this observation as the PEC for this line would be considerably smaller due to the cooler temperature (cf. Figure \ref{fig:colradpy:full_spec_many_conditions_landscape}).

Finally, \cite{gillanders2024modelling} also highlights a feature at $\sim 1.2\mu$m.
They propose both Te {\sc ii} $^3$D$_{5/2}^{\mathrm{o}}\to$ $^2$P$_{1/2}^{\mathrm{o}}$ ($\lambda = 1.23$ $\mu$m) and Te {\sc iii} $^3$P$_0\to$ $^3$P$_2$ ($\lambda = 1.22$ $\mu$m) groundstate fine-structure lines as possible candidates for this feature. This feature fades quickly and is difficult to quantify, but to be apparent in the spectra a rough luminosity of $\sim 10^{39}$ erg s$^{-1}$ is required. Based on the model presented here, these ions could contribute here but are unlikely to be the dominating factor, as to produce this order of luminosity with these lines would require $\sim 100$ $\text{M}_{\odot}$ of Te {\sc ii} or $\sim 0.1$ $\text{M}_{\odot}$ of Te {\sc iii}. This was calculated with $T_e = 0.27$ eV and $n_e = 10^7$ cm$^{-3}$ to roughly account for the +4.4d conditions \citep{gillanders2022}.

The $2.1\mu$m feature also appears in the GRB230307A spectrum \cite{levan2024heavy}. This has again been attributed to Te {\sc iii} \citep{gillanders2023heavy,levan2024heavy}, which remains plausible based on the collisional data presented here. With an integrated luminosity of $3 \times 10^{38}$ erg s$^{-1}$, the data of \cite{madonna2018neutron} was used by \cite{levan2024heavy} to esimate a mass of roughly $10^{-3} M_{\odot}$ at $T_e = 0.26$ eV and $n_e = 3\times 10^{5}$ cm$^{-3}$. The current atomic data leads to a similar required mass of $2.1 \times 10^{-3} \text{M}_{\odot}$ which is again consistent with the difference in $\Upsilon$ values. 

The GRB202307A observation is also suggestive of excess emission around 4.5$\mu$m.  \cite{gillanders2023heavy} and \cite{levan2024heavy} attribute a spectral feature due to this, with a preferred identification of Se {\sc iii} or W {\sc iii} by \cite{levan2024heavy} in agreement with the previous assignment to the Spitzer observation of AT2017gfo by \cite{Hotokezaka2022WSe}. The collisional data here requires a mass of Te {\sc ii} of $\sim 0.3$ M$_\odot$ to produce the integrated luminosity of $\sim 7\times10^{37}$ erg s$^{-1}$ \citep{gillanders2023heavy}. This is clearly an unreasonable requirement and the proposition of Te {\sc ii} contributing significantly to flux here can likely be ruled out. The assignment of Se {\sc iii} or W {\sc iii}  by \cite{Hotokezaka2022WSe} here is perhaps more promising. Given this electron temperature of 3000 K is perhaps an overestimate, it is noteworthy that the required mass of Te {\sc ii} will increase significantly with decreasing electron temperature estimates.

A caveat in making such crude mass estimates is the neglect of opacity effects. It was found for Te {\sc i-iii} in this analysis that even at late times the resonance lines (e.g the 5s$^2$ 5p$^3$ 6s $^5$S$_2^\mathrm{o}$ $\to$ 5s$^2$ 5p$^4$  $^3$P$_2$ line of Te {\sc i}) exhibited a Sobolov depth $\tau_s >> 1$. Given that the upper energies of the dipoles are likely not thermally accessible at these temperatures, opacity will likely not cause a large effect. By contrast the forbidden lines of Te {\sc i} and {\sc  ii} were found to have $\tau_s << 1$. Interestingly, the $2.1 \mu$m line of Te {\sc iii} was found to have a depth of $\tau \sim 0.5$ at 10.4d, showing marginal optical depth. This would suggest that a moderate correction of the emission for this line may arise from opacity effects, although we note that, as a contribution to the uncertainty, this is likely to be outweighed by the larger uncertainties present in electron density and temperature. Nonetheless, this emphasises that it is important that finite opacity effect are taken into account in more detailed radiative transfer modelling such as \citep{PognanNLTE,gillanders2022}.

In summary, our NLTE analysis suggests that the 2.1$\mu$m feature in both AT2017gfo can GRB230307A can indeed be explained by a significant contribution of Te {\sc iii} emission for reasonable temperatures, densities and ion masses, as previously proposed by \cite{hotokezaka2023tellurium,gillanders2024modelling} and \cite{levan2024heavy,gillanders2023heavy}. However, we find that other identifications with Te {\sc i} - {\sc iii} are not probable.

\setlength{\tabcolsep}{4pt}

\setlength{\tabcolsep}{8pt}

\begin{table*}
        \begin{tabular}{rcrrrrlcc}
            \hline
            \vspace{6mm}
            $\lambda$\phantom{00} &Index & $E_{{i}}$\phantom{00} & Lower & $E_{{j}}$\phantom{00} & Upper\phantom{-} & 
           {$A_{{j\to i}}$ } & \multicolumn{2}{c}{Luminosity $L$ (erg s$^{-1}$)}  \\  
            (nm)&(${i}$-${j}$)\phantom{0} & (cm$^{-1}$) & ${i}$\phantom{00} &   (cm$^{-1}$) &  ${j}$\phantom{00} & (s$^{-1}$) & (AT2017gfo) & (GRB230307A) \\
            \\
            \hline
            \\
            Te {\sc i} lines &&&&&& \vspace{1mm}
            \\
            2104.95&      1 -  3&          0.0&    5p$^4$ $^3$P$_{2}$&       4750.7&   5p$^4$ $^3$P$_{1}$&   2.30E+00& 4.21E+36&   8.66E+35 \\
            2124.72&      1 -  2&          0.0&    5p$^4$ $^3$P$_{2}$&       4706.5&   5p$^4$ $^3$P$_{0}$&   1.53E-02& 5.02E+35&   3.05E+35 \\
             947.16&      1 -  4&          0.0&    5p$^4$ $^3$P$_{2}$&      10557.9&   5p$^4$ $^1$D$_{2}$&   2.25E+00& 1.42E+35&   1.08E+35 \\
            1722.01&      3 -  4&       4750.7&    5p$^4$ $^3$P$_{1}$&      10557.9&   5p$^4$ $^1$D$_{2}$&   1.29E-01& 4.46E+33&   3.40E+33 \\
            1709.00&      2 -  4&       4706.5&    5p$^4$ $^3$P$_{0}$&      10557.9&   5p$^4$ $^1$D$_{2}$&   5.15E-04& 1.80E+31&   1.37E+31 \\
             542.07&      3 -  5&       4750.7&    5p$^4$ $^3$P$_{1}$&      23198.4&   5p$^4$ $^1$S$_{0}$&   2.76E+01& 2.51E+30&   9.73E+30 \\
             791.11&      4 -  5&      10557.9&    5p$^4$ $^1$D$_{2}$&      23198.4&   5p$^4$ $^1$S$_{0}$&   4.17E+00& 2.59E+29&   1.01E+30 \\
             431.06&      1 -  5&          0.0&    5p$^4$ $^3$P$_{2}$&      23198.4&   5p$^4$ $^1$S$_{0}$&   3.10E-01& 3.54E+28&   2.18E+29 \\
          226193.17&      2 -  3&       4706.5&    5p$^4$ $^3$P$_{0}$&       4750.7&   5p$^4$ $^3$P$_{1}$&   1.42E-06& 2.42E+28&   1.37E+29 \\
             225.97&      1 -  6&          0.0&    5p$^4$ $^3$P$_{2}$&      44253.0& 5p$^3$6s $^5$S$_{2}$&   7.55E+06& 1.68E+26&   4.38E+28 \\
            \\
            Te {\sc ii} lines &&&&&& \vspace{1mm}
            \\
            978.25&      1 -  2&          0.0&  5p$^3$ $^4$S$_{3/2}$&      10222.4& 5p$^3$ $^2$D$_{3/2}$&   1.72E+00&   9.84E+36&   6.97E+36\\
            805.03&      1 -  3&          0.0&  5p$^3$ $^4$S$_{3/2}$&      12421.9& 5p$^3$ $^2$D$_{5/2}$&   1.29E-01&   3.51E+35&   1.24E+36\\
           4546.55&      2 -  3&      10222.4&  5p$^3$ $^2$D$_{3/2}$&      12421.9& 5p$^3$ $^2$D$_{5/2}$&   9.27E-02&   4.46E+34&   1.58E+35\\
            486.70&      1 -  4&          0.0&  5p$^3$ $^4$S$_{3/2}$&      20546.6& 5p$^3$ $^2$P$_{1/2}$&   6.55E+00&   6.45E+33&   1.05E+34\\
            968.60&      2 -  4&      10222.4&  5p$^3$ $^2$D$_{3/2}$&      20546.6& 5p$^3$ $^2$P$_{1/2}$&   3.85E+00&   1.91E+33&   3.09E+33\\
            416.11&      1 -  5&          0.0&  5p$^3$ $^4$S$_{3/2}$&      24032.1& 5p$^3$ $^2$P$_{3/2}$&   1.07E+01&   7.78E+32&   2.54E+33\\
            724.13&      2 -  5&      10222.4&  5p$^3$ $^2$D$_{3/2}$&      24032.1& 5p$^3$ $^2$P$_{3/2}$&   1.35E+01&   5.64E+32&   1.84E+33\\
            861.31&      3 -  5&      12421.9&  5p$^3$ $^2$D$_{5/2}$&      24032.1& 5p$^3$ $^2$P$_{3/2}$&   5.62E+00&   1.98E+32&   6.44E+32\\
           1230.81&      3 -  4&      12421.9&  5p$^3$ $^2$D$_{5/2}$&      20546.6& 5p$^3$ $^2$P$_{1/2}$&   1.16E-01&   4.52E+31&   7.33E+31\\
           2869.03&      4 -  5&      20546.6&  5p$^3$ $^2$P$_{1/2}$&      24032.1& 5p$^3$ $^2$P$_{3/2}$&   2.99E-01&   3.15E+30&   1.03E+31\\
            \\
            Te {\sc iii} lines &&&&&& \vspace{1mm}
            \\
            2102.08&      1 -  2&          0.0&    5p$^2$ $^3$P$_0$ &       4757.2&   5p$^2$ $^3$P$_{1}$ & 1.83E+00&  7.21E+38&   1.44E+38\\
            2932.72&      2 -  3&       4757.2&    5p$^2$ $^3$P$_1$ &       8167.0&   5p$^2$ $^3$P$_{2}$ & 4.68E-01&  1.87E+37&   8.30E+36\\
            1224.44&      1 -  3&          0.0&    5p$^2$ $^3$P$_0$ &       8167.0&   5p$^2$ $^3$P$_{2}$ & 1.37E-02&  1.31E+36&   5.82E+35\\
             793.49&      2 -  4&       4757.2&    5p$^2$ $^3$P$_1$ &      17359.8&   5p$^2$ $^1$D$_{2}$ & 3.34E+00&  2.17E+35&   1.06E+35\\
            1087.81&      3 -  4&       8167.0&    5p$^2$ $^3$P$_2$ &      17359.8&   5p$^2$ $^1$D$_{2}$ & 3.44E+00&  1.63E+35&   7.99E+34\\
             576.04&      1 -  4&          0.0&    5p$^2$ $^3$P$_0$ &      17359.8&   5p$^2$ $^1$D$_{2}$ & 6.09E-03&  5.46E+32&   2.67E+32\\
             390.00&      2 -  5&       4757.2&    5p$^2$ $^3$P$_1$ &      30398.3&   5p$^2$ $^1$S$_{0}$ & 4.54E+01&  1.62E+31&   1.37E+32\\
             449.82&      3 -  5&       8167.0&    5p$^2$ $^3$P$_2$ &      30398.3&   5p$^2$ $^1$S$_{0}$ & 2.21E+00&  6.82E+29&   5.77E+30\\
             766.96&      4 -  5&      17359.8&    5p$^2$ $^1$D$_2$ &      30398.3&   5p$^2$ $^1$S$_{0}$ & 2.61E+00&  4.72E+29&   4.00E+30\\
             177.24&      3 -  6&       8167.0&    5p$^2$ $^3$P$_2$ &      64586.5& 5s5p$^3$ $^5$S$_{2}$ & 1.22E+06&  1.09E+21&   6.02E+25\\

            \\
           \hline
        \end{tabular}
        \caption{The ten strongest optically thin Te lines. The Luminosities are calculated for a reference mass of $10^{-3}$ M$_\odot$ for each ion stage at two sets of extrinsic conditions. The AT2017gfo is modelled with $T_e = 0.17$ eV and $n_e = 7.4\times 10^6$ cm$^{-3}$. The GRB230307A is modelled with $T_e = 0.26$ eV and $n_e = 3\times 10^5$ cm$^{-3}$. }
        \label{tab:lumo}
    \end{table*} 

\subsection{Line Ratio Diagnostics}

\begin{figure}
    \centering
    \includegraphics[width = \linewidth]{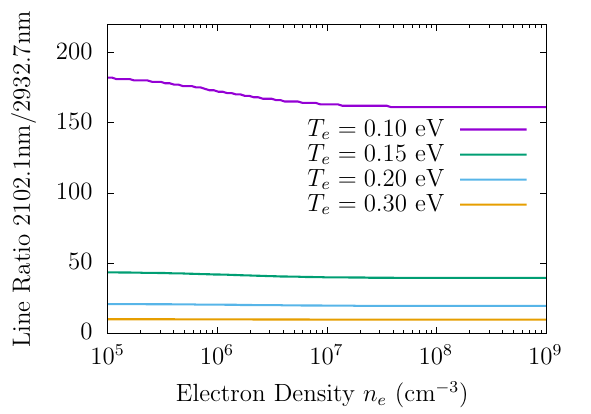}
    \caption{PEC line ratio for two Te {\sc iii} groundstate transitions as a function of eletron density at different temperatures. The transitions in question are  $^3$P$_1$ $\to$ $^3$P$_0$ ($\lambda = 2101.1$ nm) and  $^3$P$_2$ $\to$ $^3$P$_1$ ($\lambda = 2932.7$ nm). }
    \label{fig:colradpy:lr1}
\end{figure}
\begin{figure*}
    \centering
    \includegraphics[width = \linewidth]{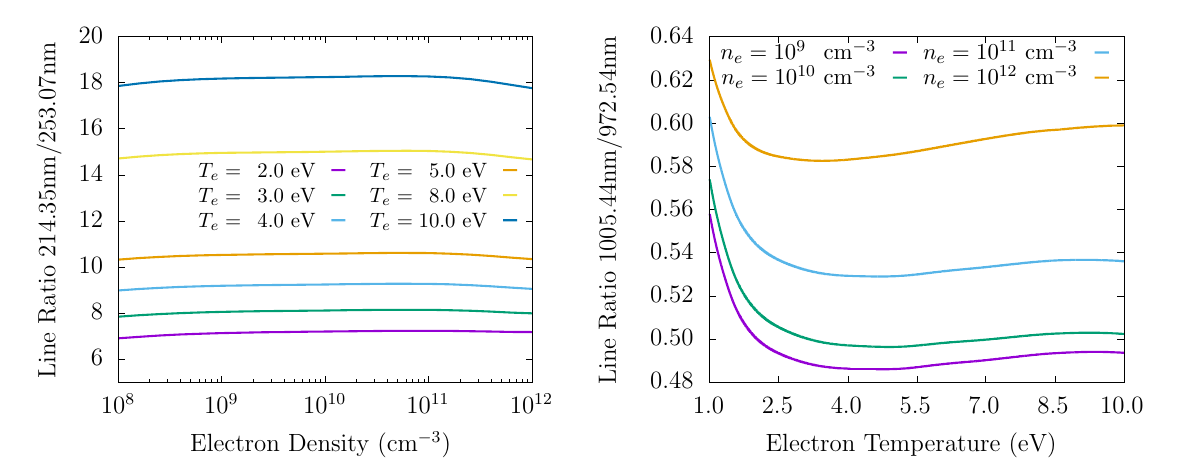}
    \caption{PEC line ratio for two Te {\sc i} transitions as a function of electron density and electron temperatures.  The line ratio between the transitions 5p$^3$6s$\phantom{}$ $^3$S$_1^\mathrm{o}$ $\to$ 5p$^4$ $^3$P$_2$ ($\lambda = 214.35$ nm) and  5p$^3$6s$\phantom{}$ $^5$S$_1^\mathrm{o}$ $\to$ 5p$^4$ $^3$P$_1$ ($\lambda = 253.07$ nm) are shown on the left panel.
    The line ratio between the transitions 5p$^3$ 6p$\phantom{}$ $^5$P$_2$ $\to$ 5p$^3$ 6s$\phantom{}$ $^5$S$_2^\mathrm{o}$ ($\lambda = 1005.42$ nm) and  5p$^3$ 6p $^5$P$_3$ $\to$ 5p$^3$ 6s $^5$S$_2^\mathrm{o}$ ($\lambda = 972.54$ nm)  are shown on the right panel as a function of temperature at densities reminiscent of those at the CTH machine \citet{bromley2020panning}. }
    \label{fig:colradpy:lr3}
\end{figure*}

The currently accepted indicative line at 2.1$\mu$m  for Te identification serves as a particularly good density diagnostic, as the upper level at 0.0434 Ry is excitable at low temperatures. While this is the case, it is in general practical to make use of line ratios. In the remainder of this section to present new possible diagnostic ratios and in particular, we present emission line ratios for the diagnosis of temperature and density. We present tools for two use cases, firstly the observation of KNe capable of observing large wavelength windows at low-medium resolution with relatively low temperatures and densities, and secondly the high resolution measurement of spectra from laboratory plasmas with shorter wavelength windows with temperatures and densities typical of that of modern plasma experiments such as those featured by \citet{bromley2020panning}.

Previous population studies have shown that the relatively low temperature of the observed KNe greatly restricts the number of observable lines, where only the first few states for Te {\sc i}-{\sc iii} can be reliably excited. Based on the spectra shown in Figure \ref{fig:colradpy:full_spec_many_conditions_landscape}, it is clear that contributions from Te {\sc i} and Te {\sc ii} are likely to be of poor use diagnostically at the low temperatures expected at the late epochs of the KNe  \citep[1700 - 2400 K $\approx$ 0.14 - 0.2 eV]{hotokezaka2023tellurium,gillanders2024modelling}. For this reason, we focus our attention on KNe diagnostics for Te {\sc iii}.  

For Te {\sc iii}, the ground state transition $^3$P$_1\to$ $^3$P$_0$ ($\lambda = 2.1$ $\mu$m) line is of particular interest. Additionally, there is a neighbouring strong transition $^3$P$_2\to$ $^3$P$_1$ ($\lambda = 2.9$ $\mu$m). \citet{hotokezaka2023tellurium} state that future observation of this line would provide conclusive evidence of Te {\sc iii}, with its ratio to the  2.1 $\mu$m line being indicative of the plasma temperature. Using the acquired collisional data, we show this line ratio as a temperature diagnostic on Figure \ref{fig:colradpy:lr1}. However, given that the $2.9$ $\mu$m line is much weaker than the 2.1 $\mu$m line at KNe conditions (see Table \ref{tab:lumo}) and therefore measurement is likely too difficult for future observation to observe this feature.

Plasmas observed at the CTH machine at Auburn University were measured to have temperatures and densities of the order of $10$ eV and $10^{12}$ cm$^{-3}$ respectively \citep{bromley2020panning}. In these relatively extreme conditions more lines become observable with a high density of lines found in the 200 - 300 nm range across the three ion stages. Of particular importance are electric dipoles at these conditions where the levels are thermally accessible. Line ratios across all three ions were searched for systematically. Many of the line ratios studied in short wavelength windows exhibited large dependence on both temperature and density. From the data presented here, it was found the best temperature and density diagnostics both came from Te {\sc i}. This was determined from the individual strength of both lines, and the dependence of the ratio on temperature and density. For a temperature diagnostic, the line ratio between the Te {\sc i} transitions 5p$^3$6s$\phantom{}$ $^3$S$_1^\mathrm{o}$ $\to$ 5p$^4$ $^3$P$_2$ ($\lambda = 214.35$ nm) and  5p$^3$6s$\phantom{}$ $^5$S$_1^\mathrm{o}$ $\to$ 5p$^4$ $^3$P$_1$ ($\lambda = 253.07$ nm) is shown on the left panel of Figure \ref{fig:colradpy:lr3}. Density diagnostics remain sparse with few potential candidates found. One promising line ratio is that between the Te {\sc i} transitions 6p$\phantom{}$ $^5$P$_2$ $\to$ 6s$\phantom{}$ $^5$S$_2^\mathrm{o}$ ($\lambda = 1005.43$ nm) and  6p $^5$P$_3$ $\to$ 6s $^5$S$_2^\mathrm{o}$ ($\lambda = 972.57$ nm), which is shown on the right panel of Figure \ref{fig:colradpy:lr3}. While there is notable temperature dependence at very low temperatures, with increasing temperature this decays and the line ratio remains a strong function of electron density allowing for a potentially accurate assessment of the electron density. It is intended for these line ratios to guide the plasma community quantitative measurements of electron temperature and density of Te plasmas.

\section{Conclusions and Outlook}\label{sec:conclusions}

In this work the Dirac-Fock method was used to produce model atomic structures for all three ions. These ab initio calculations were found to exhibit good agreement with the literature energy levels and transition probabilities. The three structure models were incorporated into the DARC-R-matrix packages to compute the collision strength profiles with a representative example of transitions shown. Maxwellian averaged collisions strengths were calculated.

Collisional radiative modelling was carried out with the ColRadPy routines \citep{johnson2019colradpy} to produce model synthetic spectra.  It was found that generally the only Te line that is strong enough to produce features in KNe with reaonsable masses was the Te {\sc iii} line at 2.1 $\mu$m. Additionally, luminosity estimates reinforce the calculations  of \cite{hotokezaka2023tellurium} and show the most prominent feature from Te is certainly the 2.1 $\mu$m line. The required mass of Te {\sc iii} is reasonable for this species to be a significant and perhaps even dominant contributor to this feature in the AT2017GF0. Based on the basic NLTE analysis here, it is clear that Te in general deviates far enough from thermal equilibrium for NLTE simulations to be necessary to appropriately model the population dynamics of the plasma in the nebular phase. Additionally, under the KNe conditions studied here we have found that features such as the 1.2 $\mu$m and 4.5 $\mu$m lines are unlikely to be contributed to by Te while simultaneously explaining the 2.1 $\mu$m line. This Te {\sc iii} line remains the prominent identifying characteristic of Te. In spite of the relative difference between the data presented here and that of \cite{madonna2018neutron}, mass estimates of Te {\sc iii} remain within the same order of magnitude with relative agreement.  Incoporating new data into early time LTE models reveals very little contribution to the spectra from Te.

Quantitative line-flux ratios have been proposed as a temperature diagnostic for the KNe plasma. At the considered KNe conditions, diagnostic ratios for density profiling were not found.  Additionally, laboratory benchmarks have been proposed in the form of line ratios for Te {\sc i} at more extreme temperatures and densities. Line ratios in Te {\sc ii} and {\sc iii} were found to exhibibit strong behaviour on both temperature and density for lines of interest for modern plasmas experiments.

A key limiting factor in modelling the KNe and exploiting expanding atomic datasets is the understanding of the ionization states of the atomic species in the plasma. With further calculation of ionization and recombination rates, the restriction to ion-masses above could potentially be relaxed and element masses can be more quantitatively addressed.  This will be further aided by future work where, electron-impact-ionization and photoionization cross sections and recombination rates will be calculated. The data presented in this work should prove beneficial for more sophisticated NLTE KNe simulations. This combined with the present analysis will potentially aid the characterisation of the currently observed and any future KNe spectra.

\section*{Funding}

Funded/Co-funded by the European Union (ERC, HEAVYMETAL, 101071865). Views and opinions expressed are however those of the author(s) only and do not necessarily reflect those of the European Union or the European Research Council. Neither the European Union nor the granting authority can be held responsible for them.

\section*{Acknowledgements}\label{sec:acknowledgements}

We thank our colleagues at Queen's University Belfast and Auburn University for helpful discussion. We thank Albert Sneppen for reading the manuscript and providing useful comments. We are grateful for use of the computing resources from the Northern Ireland High Performance Computing (NI-HPC) service funded by EPSRC (EP/T022175). This research made use of \textsc{tardis}, a community-developed software package for spectral synthesis in supernovae \citep{KerzendorfSimTardis, vogl2019,kerzendorf_2023_8244935}. The development of \textsc{tardis} received support from the Google Summer of Code initiative and from ESA’s Summer of Code in Space program. \textsc{tardis} makes extensive use of Astropy and PyNE.

\section*{Data Availability}

The effective collision strengths for Te {\sc i}-{\sc iii} are recorded in the standard adf04 format on both this articles online supplementary material and at http://open.adas.ac.uk/. Other data underlying this article will be shared on reasonable request to the corresponding author. 



\bibliographystyle{mnras}
\bibliography{bibliography}§ 





\bsp	
\label{lastpage}
\end{document}